\documentclass{JHEP3}    
\usepackage{epsfig}

\newcommand{\eref}[1]{(\ref{#1})}

\newcommand{\fref}[1]{Figure~\ref{#1}}
\newcommand{\cref}[1]{Chapter~\ref{#1}}
\newcommand{\beq}{\begin{equation}}
\newcommand{\eeq}{\end{equation}}
\newcommand{\ba}{\begin{array}}
\newcommand{\ea}{\end{array}}
\newcommand{\bcenter}{\begin{center}}
\newcommand{\ecenter}{\end{center}}

\def\IB{\relax\hbox{$\inbar\kern-.3em{\rm B}$}}
\def\IC{\relax\hbox{$\inbar\kern-.3em{\rm C}$}}
\def\ID{\relax\hbox{$\inbar\kern-.3em{\rm D}$}}
\def\IE{\relax\hbox{$\inbar\kern-.3em{\rm E}$}}
\def\IF{\relax\hbox{$\inbar\kern-.3em{\rm F}$}}
\def\IG{\relax\hbox{$\inbar\kern-.3em{\rm G}$}}
\def\IGa{\relax\hbox{${\rm I}\kern-.18em\Gamma$}}
\def\IH{\relax{\rm I\kern-.18em H}}
\def\IK{\relax{\rm I\kern-.18em K}}
\def\IL{\relax{\rm I\kern-.18em L}}
\def\IP{\relax{\rm I\kern-.18em P}}
\def\IR{\relax{\rm I\kern-.18em R}}
\def\IZ{\relax\ifmmode\mathchoice
{\hbox{\cmss Z\kern-.4em Z}}{\hbox{\cmss Z\kern-.4em Z}}
{\lower.9pt\hbox{\cmsss Z\kern-.4em Z}}
{\lower1.2pt\hbox{\cmsss Z\kern-.4em Z}}\else{\cmss Z\kern-.4em Z}\fi}
\def\II{\relax{\rm I\kern-.18em I}}

\def\sCC{{\kern 0.27em\vrule height1.45ex width0.03em depth0em
          \kern-0.30em\rm C}}
\def\C{{\mathchoice
  {\sCC}
  {\sCC}
  {\kern 0.225em \vrule height1.05ex width0.025em depth0em \kern-0.25em \rm C}
  {\kern 0.180em \vrule height0.78ex width0.02em depth0em \kern-0.2em \rm C}
        }}
\def\sHH{{\rm I\kern-.16em{}H}}
\def\H{{\mathchoice
  {\sHH}
  {\sHH}
  {\rm I\kern-.13em{}H}
  {\rm I\kern-.13em{}H} }}
\def\sNN{{\rm I\kern-.16em{}N}}
\def\N{{\mathchoice
  {\sNN}
  {\sNN}
  {\rm I\kern-.12em{}N}
  {\rm I\kern-.10em{}N} }}
\def\sPP{{\rm I\kern-.16em{}P}}
\def\P{{\mathchoice
  {\sPP}
  {\sPP}
  {\rm I\kern-.12em{}P}
  {\rm I\kern-.10em{}P} }}
\def\sQQ{{\kern 0.27em \vrule height1.45ex width0.03em depth0em
          \kern-0.30em \rm Q}}
\def\Q{{\mathchoice
        {\sQQ}
        {\sQQ}
  {\kern 0.225em \vrule height1.05ex width0.025em depth0em \kern-0.25em \rm Q}
  {\kern 0.180em \vrule height0.78ex width0.020em depth0em \kern-0.20em \rm Q}
        }}
\def\sRR{{\rm I\kern-0.16em{}R}}
\def\R{{\mathchoice
  {\sRR}
  {\sRR}
  {\rm I\kern-0.12em{}R}
  {\rm I\kern-0.10em{}R} }}
\def\sZZ{{\rm Z\kern-0.32em{}Z}}
\def\Z{{\mathchoice
  {\sZZ}
  {\sZZ} 
  {\rm Z\kern-0.3em{}Z}     
  {\rm Z\kern-0.25em{}Z} }}  
\def\ZZZ{{\rm Z\kern-0.24em{}Z}}
\def\sII{{\rm I\kern-0.16em{}I}}
\def\I{{\mathchoice
  {\sII}
  {\sII}
  {\rm I\kern-0.12em{}I}
  {\rm I\kern-0.10em{}I} }}

\def\inbar{\,\vrule height1.5ex width.4pt depth0pt}
\font\cmss=cmss10 \font\cmsss=cmss10 at 7pt

\def\smiley{\hbox{\large$\bigcirc$\hspace{-0.80em}\raise.2ex
\hbox{$\cdot\cdot$}\kern-.61em\lower.2ex\hbox{\scriptsize$\smile$}}\ }
\def\frowny{\hbox{\large$\bigcirc$\hspace{-0.80em}\raise.2ex
\hbox{$\cdot\cdot$}\kern-.635em\lower.2ex\hbox{\scriptsize$\frown$}}\ }

\def\I{{\rlap{1} \hskip 1.6pt \hbox{1}}}

\makeatletter
\let\hangafter\@hangfrom
\makeatother

\newcommand{\drawsquare}[2]{\hbox{%
\rule{#2pt}{#1pt}\hskip-#2pt
\rule{#1pt}{#2pt}\hskip-#1pt
\rule[#1pt]{#1pt}{#2pt}}\rule[#1pt]{#2pt}{#2pt}\hskip-#2pt
\rule{#2pt}{#1pt}}

\newcommand{\fund}{\raisebox{-.5pt}{\drawsquare{6.5}{0.4}}}
\newcommand{\antifund}{\overline{\fund}}

\newcommand{\be}{\begin{equation}}
\newcommand{\ee}{\end{equation}}
\newcommand{\bea}{\begin{eqnarray}}
\newcommand{\eea}{\end{eqnarray}}
\newcommand{\bean}{\begin{eqnarray*}}
\newcommand{\eean}{\end{eqnarray*}}

\newcommand{\beqa}{\begin{eqnarray}}
\newcommand{\eeqa}{\end{eqnarray}}
\newcommand{\id}{{\bf 1}}
\def\ov{\overline}

\def\diag{{\rm diag \,}}

\preprint{
CERN-PH-TH/2005-007\\
IFT-UAM/CSIC-04-64\\ 
MIT-CTP -3583\\ 
} 

\title{Multi-Flux Warped Throats and Cascading Gauge Theories}

\author{Sebastian Franco$^\dagger$, Amihay Hanany$^\dagger$ and Angel M. Uranga$^\#$
\footnote{
Research supported in part by the CTP and the LNS
of MIT and the U.S. Department of Energy under cooperative agreement
$\#$DE-FC02-94ER40818, the BSF American--Israeli Bi--National Science Foundation and 
the CICYT, Spain, under project FPA2003-02877. 
A. H. is also supported by  a DOE OJI award.}
\\
~\\
$^\dagger$Center for Theoretical Physics, Massachusetts Institute of Technology, \\
Cambridge, MA 02139, USA.\\
\email{sfranco, hanany@mit.edu} \\
$^\#$Instituto de F\'{\i}sica Te\'orica, Facultad de Ciencias, C- XVI \\
Universidad Aut\'onoma de Madrid, 28049 Madrid, Spain\\
and TH Division, CERN, CH-1211 Geneve 23, Switzerland\\
\email{angel.uranga@uam.es, angel.uranga@cern.ch}
}

\abstract{We describe duality cascades and their infrared behavior for
systems of D3-branes at singularities given by complex cones over del 
Pezzo surfaces (and related examples), in the presence of fractional branes. 
From the gauge field theory viewpoint, we show that D3-branes probing the 
infrared theory have a quantum deformed moduli space, given by a complex 
deformation of the initial geometry to a simpler one. This implies 
that for the dual supergravity viewpoint, the gauge theory strong 
infrared dynamics smoothes out the naked singularities of the recently 
constructed warped throat solutions with 3-form fluxes, describing the 
cascading RG flow of the gauge theory. This behavior thus generalizes 
the Klebanov-Strassler deformation of the conifold. We describe several 
explicit examples, including models with several scales of strong gauge 
dynamics. In the regime of widely separated scales, the dual supergravity 
solutions should correspond to throats with several radial regions with 
different exponential warp factors. These rich throat geometries are expected
to have interesting applications in compactification and model building. 
Along our studies, we also construct explicit duality cascades for gauge 
theories with irrational R-charges, obtained from D-branes probing complex 
cones over $dP_1$ and $dP_2$.
}

\begin{document}

\section{Introduction}

Much insight into the gauge/gravity correspondence has been obtained from 
the study of D3-branes at singularities. In the simplest situations where 
only regular D3-branes are present, the resulting gauge theories are 
conformal, and are dual to superstring backgrounds of the form 
AdS$_5\times X_5$, where $X_5$ is the base of the real cone describing the 
singular manifold \cite{Klebanov:1998hh}. This has led to important 
extensions of the AdS/CFT correspondence to situations with reduced 
supersymmetry.

Conformal invariance can be broken by adding fractional D3-branes (e.g. 
D5-branes wrapped over collapsed 2-cycles at the tip of the singularity).
The resulting renormalization group (RG) flow sometimes takes the form of 
a {\em duality cascade}. In a duality cascade, Seiberg duality is used to 
change to a dual description every time any of the gauge groups goes 
to infinite coupling. The idea of a cascading RG flow was first 
introduced in \cite{Klebanov:2000hb}, for the gauge theory on D-branes 
over a conifold singularity.

The ultraviolet (UV) behavior of cascading theories is markedly different 
from that of ordinary field theories. Instead of having a UV fixed point, 
they have an infinite tower of dual theories with a steadily increasing 
number of colors and matter fields towards the UV. This increase can
sometimes be linear as in \cite{Klebanov:2000hb}, or can be much faster,
with a power law or even exponential behavior. In the latter cases, the dualization
scales generally present a UV accumulation point, leading to a duality wall 
\cite{Hanany:2003xh,Franco:2003ja}.

A supergravity solution describing the UV region of the conifold cascade 
was found by Klebanov and Tseytlin (KT) in \cite{Klebanov:2000nc}. This 
solution is well behaved at large energies but has a naked singularity
in the infrared (IR). A full solution, which asymptotes the one of KT at 
large energies but is well behaved in the IR was later presented by 
Klebanov and Strassler (KS) in \cite{Klebanov:2000hb}. Instead of being 
based on the singular conifold, it is constructed using the {\em deformed 
conifold}. The 3-cycle inside the deformed conifold remains of finite 
size in the IR, avoiding the singular behavior. On the gauge theory side, 
the IR singularity is eliminated by strong coupling effects, whose scale 
is related to the dual 3-cycle size.

In \cite{Franco:2004jz}, UV solutions, similar to that of KT, were 
constructed for complex cones over del Pezzo surfaces $dP_n$, for $3\leq n \leq 8$. 
These supergravity 
solutions also suffer from the same problems in the IR. Contrary to what 
happens for the conifold, explicit metrics describing either the 
non-spherical horizons or their deformations are not known. Therefore it 
remains an open question to develop methods to understand the infrared 
behavior of these theories in their dual versions. The purpose 
of this paper is to use the strong coupling dynamics of the dual gauge 
theories to extract as much information as possible regarding these 
deformations. In particular we will show a precise agreement between the 
field theory analysis of D3-branes probing the infrared of the cascades 
and the complex deformations of the initial geometries. This strongly 
supports the existence of completely smooth supergravity descriptions of 
the complete RG flow for (some of) these non-conformal gauge theories.
In addition, our techniques are valid for other geometries, suggesting the 
existence of cascades and infrared deformations for other quiver gauge 
theories.

Although our examples are analogous to the conifold in some respects, 
the gauge theories and corresponding geometries are notably richer in 
others. For instance, we will encounter that these gauge theories 
generically give rise to several dynamical scales. In the regime of 
widely separated scales, the flow among these scales is to a great 
approximation logarithmic. The supergravity duals thus correspond to 
logarithmic throats with different warp factors, patched together at some 
transition scales. Clearly these topologically richer throats deserve
further study.

Before proceeding, it is important to point out that our analysis shows 
that a smoothing of the singularity by a complex deformation may not be 
possible for some geometries, or even for all possible assignments of 
fractional branes in a geometry. Our methods give a clear prescription for 
when this is the case. A class of examples of this kind is provided by the
countable infinite family of 5d horizons with $S^2 \times S^3$ topology, 
for which explicit metrics have been constructed in 
\cite{Gauntlett:2004zh,Gauntlett:2004yd,Gauntlett:2004hh,Gauntlett:2004hs,Martelli:2004wu}. These geometries 
are labeled by two positive integers $p>q$ and are denoted $Y^{p,q}$. In 
\cite{Benvenuti:2004dy}, the quiver theories living on the world-volume of 
D3-branes probing metric cones over $Y^{p,q}$ geometries were derived. Impressive
checks of the AdS/CFT correspondence for these models, such as matching the field theory R-charges and
central charge $a=c$ with the corresponding geometric computations were carried out
in full generality \cite{Benvenuti:2004dy}. 
Recently, warped throat supergravity solutions dual to cascades in the 
$Y^{p,q}$ quivers were constructed in \cite{Ejaz:2004tr}. These solutions 
exhibit a naked singularity, and we show that for the particular subclass 
of $Y^{p,0}$ a complex deformation removes the IR singularity. However, 
in the general case these geometries do not admit complex deformations to 
smooth out their infrared behavior. It would be interesting to understand 
such examples, and we leave this question for future research.

\medskip

The paper is organized as follows. In Section \ref{cascading} we provide 
some background material. In section \ref{revconifold} we review the KS 
conifold. In section \ref{sugrathroats} we describe the supergravity 
throats constructed in \cite{Franco:2004jz}, and generalizations. 
In sections \ref{deformedgeom} and \ref{topocons} we present a framework to 
determine possible geometric deformations for general local Calabi-Yau geometries 
using $(p,q)$ web diagrams, and discuss a 
topological property of the corresponding RG flow solutions. In section 
\ref{generaldeform} we introduce our approach to show that the strong 
gauge theory dynamics induces the complex deformation of the initial 
geometries to simpler ones. These arise as quantum deformations of the 
moduli space of the gauge theory describing D3-branes probing the infrared 
dynamics.

In Section \ref{warmup} we describe some simple examples of RG cascading 
flows and infrared deformations, in several cases with a single strong 
dynamics scale. The examples include the cone over $F_0$, the cone over 
$dP_2$, and the suspended pinch point (SPP) singularity. In 
subsequent sections we present examples with several strong 
dynamics scales. In Section \ref{dpthree} we study the case of the cone over 
$dP_3$, which admits a two-scale deformation following the pattern 
$dP_3\to$ conifold $\to$ smooth ($\IC^3$). In Section \ref{further} we present 
further two-scale examples, namely $dP_4\to$ SPP $\to$ smooth, and 
$PdP_3 \to \IC^2/\IZ_2\to$ enhan\c{c}on.

Section \ref{conclusions} contains our concluding remarks. Appendix 
\ref{mesonic} presents an alternative approach for the field theory analysis of the 
mesonic branch, while Appendix \ref{toric} provides a detailed 
description of the deformations in toric geometry. Finally Appendix \ref{yp0} 
describes the field theory description of the smoothing for real 
cones over the $Y^{p,0}$ manifolds.

\section{Cascading throats}
\label{cascading}

In this section we lay out our approach to cascading RG flows. We first 
discuss the supergravity duals that describe logarithmic flows, beginning 
with a review of the well known conifold example and then moving on to 
generalizations to other geometries. We then explain how to identify 
extremal transitions using $(p,q)$ web diagrams. Finally, 
we discuss how these geometric deformations are generated by the strong coupling dynamics 
of the gauge theory.

\subsection{Review of the conifold}
\label{revconifold}

To frame the forthcoming discussion, it is convenient to review the case 
of the conifold. The 
${\cal N}=1$ supersymmetric gauge theory on $N$ D3-branes at a conifold 
singularity 
\cite{Klebanov:1998hh}, in the 
presence of $M$ fractional branes (i.e. D5-branes wrapped over the 2-cycle in the base of the conifold), 
is given by a gauge group $SU(N)\times SU(N+M)$, with two 
chiral multiplets $A_1$, $A_2$ in the representation $(\fund,\antifund)$ and two multiplets $B_1$, $B_2$ in the representation
$(\antifund,\fund)$. The superpotential is $W=A_1B_1A_2B_2-A_1B_2A_2B_1$. In order to keep the notation short, we 
leave the superpotential couplings and the trace over color indices implicit. We will adopt this 
convention when presenting all the forthcoming superpotentials.

As discussed in \cite{Klebanov:2000hb}, for $M\ll N$ the theory undergoes a duality cascade as it flows to the 
infrared, at each step of which the highest rank gauge group becomes strongly coupled and is replaced by 
its Seiberg dual. Since the gauge theory of the conifold is self-dual (up to a modification of the gauge group ranks),\footnote{Strictly speaking the term self-dual is exact at the conformal point for $M=0$. This notion of self-duality is then borrowed to the case $M\not=0$, where the superpotential stays the same while the ranks are changed.} the cascade is fully specified by the sequence of gauge groups
\beqa
SU(N)\times SU(N+M) \rightarrow SU(N)\times SU(N-M) \rightarrow SU(N-2M)
\times SU(N-M)\rightarrow \nonumber
\ldots
\eeqa
which shows that the number of effective D3-branes decreases along the flow, while the number of fractional
branes, given by the difference in ranks between the two gauge groups, remains constant and equal to $M$.
The cascade proceeds until this number is comparable with $M$. For $N$ a multiple of $M$, the 
infrared theory has chiral symmetry breaking and confinement, and shares 
some features with
${\cal N}=1$ $SU(M)$ SYM. Besides the heuristic field theory arguments, this picture is strongly 
supported by the dual supergravity solutions, which we now turn to describe.

In the absence of fractional branes the gauge theory on D3-branes at a conifold singularity is superconformal, 
and its supergravity dual is given by Type IIB theory on AdS$_5\times T^{1,1}$. The 5-manifold 
$T^{1,1}$ is topologically $S^2\times S^3$, and may be regarded as an $S^1$ fibration over 
$S^2\times S^2$. Denoting $\sigma_i$ the 2-forms dual to the two $S^2$'s, we define for future 
convenience the K\"ahler class $\omega=\sigma_1+\sigma_2$ and the orthogonal combination 
$\phi=\sigma_1-\sigma_2$.

In the presence of $M$ fractional branes, conformal invariance of the gauge theory is broken, and 
the supergravity dual is no longer AdS$_5\times T^{1,1}$. In the UV, the supergravity dual is 
a particular case of the throats to be described in section \ref{sugrathroats}. Sketchily, it is a
warped version of AdS$_5\times T^{1,1}$, with warping sourced by non-trivial RR and NSNS 
3-form fluxes supported on $\phi$, 
\beq
G_3 = F_3 - \frac{i}{g_s} H_3= M (\eta + i \frac{dr}{r}) \wedge \phi
\eeq
where $\eta$ is a 1-form along the $S^1$ fiber in $T^{1,1}$, and $r$ is the radial coordinate. 
In intuitive terms, the RR flux (related to $F_3$) is 
sourced by the fractional branes in the dual description, while the NSNS flux (related to $H_3$) leads to a 
logarithmic running of the relative inverse squared gauge coupling of the field theory. The fluxes 
also lead to a radially varying integral of the 5-form over $T^{1,1}$, which reproduces the decrease in the 
number of D3-branes in the duality cascade of the field theory. The solution does not contain, even
asymptotically, an AdS$_5$. This is accordance with the fact that the gauge theory does not have a conformal
fixed point in the presence of fractional branes.

The above solution, first studied in \cite{Klebanov:2000nc}, if extended to the IR, leads to a 
naked singularity. Intuitively, this is because the above supergravity description misses the 
strong coupling dynamics taking place near the end of the cascade. The full solution in 
\cite{Klebanov:2000hb} is smooth, due to a non-trivial modification of the above ansatz in the 
infrared. In the IR, the geometry is a deformed conifold, and has a finite size $S^3$, which 
supports the RR 3-form flux. The size of this 3-cycle is related to the scale of strong dynamics 
of the dual gauge theory. The complete solution is a warped deformed conifold, with imaginary 
self-dual 3-form fluxes which are moreover $(2,1)$-forms and thus preserve 
supersymmetry \cite{Gubser:2000vg,Grana:2000jj}. In the UV, the  
full solution asymptotes the warped version of AdS$_5\times T^{1,1}$ described above, while in 
the IR it contains a non-trivial 3-cycle supporting the flux.

Overall, the gauge/gravity correspondence is a relation between the field theory, described by 
fractional D3-branes on (i.e. D5-branes on the 2-cycle of) a resolved 
conifold, and the supergravity solution, described by 3-form fluxes on a 
deformed conifold. Namely a brane-flux transition taking place between two geometries related
by an extremal transition where a 2-cycle disappears and is replaced by a 3-cycle 
\cite{Vafa:2000wi,Cachazo:2001jy,Cachazo:2001gh,Cachazo:2001sg}. In our case the geometries
under consideration are toric, and can be visualized using web or toric diagrams 
\cite{Aharony:1997ju,Aharony:1997bh}. The geometric interpretation
of webs is discussed in detail in \cite{Leung:1997tw}. In these pictures, 
finite segments and faces of the initial web correspond to 2- and 4-cycles 
in the resolution phase, while 3-cycles correspond to segments joining 
the different sub-webs in the deformation phase.
The geometrical  
transition is nicely depicted using web diagrams for the conifold 
geometry, as shown in Figure \ref{conitrans}. The detailed geometric 
description of the deformation is described in Appendix \ref{toric}.

\begin{figure}[ht]
  \epsfxsize = 14cm
  \centerline{\epsfbox{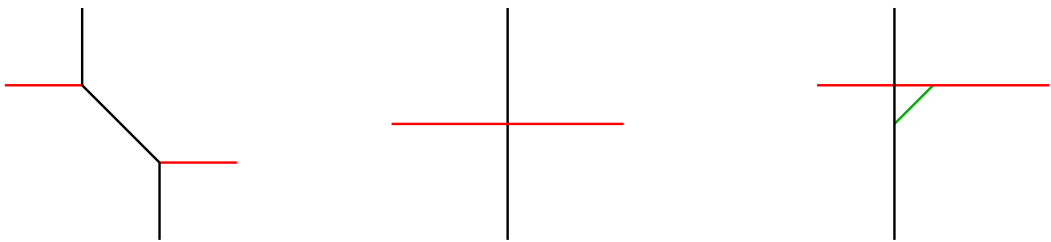}}
  \caption{Conifold extremal transition. The finite segment in the first figure represents an
           $S^2$, with an area proportional to the length of the segment, while the green segment in the last figure corresponds to an $S^3$ with a volume proportional to the distance between the two infinite lines.}
  \label{conitrans}
\end{figure}

For future convenience, it is useful to review the matching between the deformation of the 
geometry and the infrared dynamics of the field theory. In particular, and following 
\cite{Klebanov:2000hb},
we may recover the deformed conifold geometry as the moduli space of 
D3-branes probing the infrared end of the cascade. 

A simple derivation follows by considering the infrared theory in the presence of $M$ additional D3-branes.
This is described by a conifold 
gauge theory with gauge group $SU(2M)\times SU(M)$, with the chiral multiplets $A_r$, $B_r$, 
$r=1,2$ and superpotential $W=A_1B_1A_2B_2-A_1B_2A_2B_1$. The non-perturbative dynamics may be 
determined by assuming, to begin with, that the $SU(M)$ gauge factor is weakly coupled and acts as 
a spectator, corresponding to a global flavor symmetry. Then the gauge factor $SU(2M)$ 
has $N_f=N_c$ and develops a quantum deformation of its moduli space. Introducing the 
four mesons $M_{rs}=A_rB_s$, which transform in the adjoint of $SU(M)$, and the baryons ${\mathcal 
C}$, ${\tilde {\mathcal C}}$, the quantum modified moduli space is described by
\beqa
\det(M_{11})\det(M_{22})-\det(M_{12})\det(M_{21})-{\mathcal C}{\tilde {\mathcal C}}=\Lambda^{4M},
\eeqa
where $\Lambda$ is the dynamical scale of the $SU(2M)$ gauge theory.
The constraint may be implemented in the superpotential by introducing a Lagrange multiplier chiral field
$X$, so it reads
\beqa
W=M_{11}M_{22}-M_{12}M_{21}- X(\det {\cal M}-{\mathcal C}{\tilde {\mathcal C}}-\Lambda^{4M}),
\eeqa
with $\cal M$ a $2M\times 2M$ matrix whose blocks are the $M\times M$ matrices, 
$M_{11},M_{12},M_{21},M_{22}$.
The quantum constraint forces some of the mesons or baryons to acquire vevs. 
As discussed in \cite{Klebanov:2000hb}, the dynamics of the probes is 
obtained along the mesonic branch\footnote{\label{baryon} As mentioned already in \cite{Klebanov:2000hb}, 
the baryonic 
branch describes instead the continuation of the cascade down to the 
endpoint $SU(M)$ theory. That the infrared theory at the end of the 
cascade is in the baryonic branch is supported by the identification 
in the supergravity solution of the Goldstone mode associated to the 
spontaneous breaking of baryon number symmetry, and the identification of 
the D1-brane as an axionic string \cite{Gubser:2004qj,Schvellinger:2004am,Butti:2004pk}.}, which 
corresponds to 
\beqa
X=\Lambda^{4-4M} \quad ; \quad {\mathcal C}={\tilde {\mathcal C}}=0 \quad 
;\quad 
\det{\cal M}=\Lambda^{4M}
\eeqa

The vacuum is parametrized by  the vevs of the mesons $M_{ij}$, subject to the quantum 
constraint. 
This can be seen to correspond to $M$ D3-brane probes moving in a deformed conifold. To make this 
more manifest and to simplify the discussion, it is convenient to restrict to the Abelian case. 
This is sensible, because all the information about the non-Abelian gauge dynamics has been
already included, and because we are not turning on baryonic degrees of 
freedom.\footnote{\label{U1issue}A more precise statement would be to 
stick to the non-Abelian case, without overall $U(1)$, but study the dynamics 
along the generic mesonic Higgs branch. Our results below would arise for 
the relative $U(1)$'s controlling the relative positions of the D-branes. The trick of simplifying the 
discussion by restricting to the Abelian quiver theory is a standard manipulation for branes at 
singularities, see \cite{Morrison:1998cs} for further discussion.} The moduli space of the single 
D3-brane probe in this case is 
\beqa
M_{11}M_{22}-M_{12}M_{21}=\Lambda^4
\eeqa
namely, a deformed conifold geometry.
Hence the strong coupling dynamics of the field theory encodes the deformed 
geometry at the infrared end of the cascade, dictating the size of the finite
$S^3$.

The general idea is that the gauge theory living on the D-brane world volume perceives
the deformed geometry that becomes important at a given scale as a quantum deformation
of its moduli space. This technique will generalize to more complicated cascades and infrared 
behaviors in the next sections.

\subsection{The supergravity throats}
\label{sugrathroats}

As we discussed, the main support for the idea of a cascading RG flow for the conifold 
comes from the supergravity dual description \cite{Klebanov:2000nc}. In \cite{Franco:2004jz}, 
analog supergravity solutions were constructed for del Pezzo surfaces. In this section we 
review such solutions and discuss the possibility of extending the ideas in \cite{Franco:2004jz}
to other geometries.

These solutions are concrete examples of the Type IIB supergravity solutions introduced by Gra\~{n}a and 
Polchinski in \cite{Grana:2000jj} (see \cite{Dasgupta:1999ss,Giddings:2001yu} for related backgrounds in compactifications). 
The starting point is a warped product of four dimensional Minkowski space and a Calabi-Yau 3-fold $X$

\beq
ds^2 = Z^{-1/2} \eta_{\mu\nu} dx^\mu dx^\nu + Z^{1/2} ds^2_X 
\label{metric_warped}
\eeq
with the warp factor $Z$ depending only on internal coordinates of the Calabi-Yau.
In addition there is a 3-form flux

\beq
G_3 = F_3 - \frac{i}{g_s} H_3
\label{G3_1}
\eeq

It was shown in \cite{Grana:2000jj} that \eref{metric_warped} and \eref{G3_1} lead
to a solution that preserves $\mathcal{N}=1$ supersymmetry provided that $G_3$
has support only on the Calabi-Yau $X$, is imaginary self-dual with respect to
the Hodge star on $X$, and is a harmonic $(2,1)$ form.

More specifically, we will be interested in cases in which the Calabi-Yau is a complex cone over a del Pezzo surface. Thus, 
its metric has the typical form

\beq
ds_X^2 = dr^2 + r^2 \eta^2+ r^2 h_{a \bar b} dz^a d{\bar z}^{\bar b} 
\eeq  
where $\eta = \left(\frac{1}{3} d\psi + \sigma\right)$ and $h_{a \bar b}$
denotes the K\"{a}hler-Einstein metric on the del Pezzo surface.
It is important to remember that $dP_n$ only admits K\"{a}hler-Einstein metrics
for $n \geq 3$, and thus our construction will be valid in this range and will not
be applicable to the first del Pezzos.\footnote{$dP_0$ also admits a K\"ahler Einstein metric but 
cannot admit fractional D3-branes, due to the absence of collapsing 2-cycles, and therefore will not 
be considered here.} One however expects that gauge 
theory cascades for these theories exist, and presumably correspond to other throat structures (indeed 
the supergravity dual of a $dP_1$ cascade belongs to the class recently 
constructed in \cite{Ejaz:2004tr}).

For $dP_n$, $h^{2,0}=0$ and $h^{1,1} = n+1$. Thus, there are $n+1$ harmonic
$(1,1)$ forms. One of them is the self-dual K\"{a}hler form $\omega$. It is possible
to pick the remaining $(1,1)$ forms $\phi_I$, $I=1 \ldots n$, such that 

\beq
\phi_I \wedge \omega = 0 
\eeq

Also, these forms are anti-selfdual. With this basis at hand, it is straightforward to construct the 3-form flux

\beq
G_3 = \sum_{I=1}^k a^I (\eta + i \frac{dr}{r}) \wedge \phi_I 
\label{G3_2}
\eeq
where, at this point, the $a_I$ are constant coefficients determining the 
solution. It is easy to check that \eref{G3_2} satisfies all the conditions presented above 
and leads to a supersymmetric solution.

The warp factor in \eref{metric_warped} becomes, for $dP_n$, 

\beq
Z(r) = \frac{2 \cdot 3^4}{9-n} \alpha'^2 g_s^2 \left(\frac{\ln(r/r_0)}{r^4} + \frac{1}{4r^4} \right) \sum_{i,j}  M^I A_{IJ} M^J
\eeq

The number of fractional branes $M^J$ is measured by the integrals of the RR 3-form $F_3$ over the 
3-cycles in the 5-dimensional base, obtained by fibering the $U(1)$ fiber over the 3-cycles dual 
to $\phi_I$ (namely, over the 3-cycles dual to $\eta\wedge \phi_I$). The solutions describe RG flows 
in which the number of D5-branes of each type remains constant
\beq
a^J = 6 \pi \alpha' M^J
\eeq
and the effective number of D3-branes runs logarithmically with the scale. 

\beq
N=\frac{3}{2\pi} g_s \ln(r/r_0)\sum_{I,J} M^I A_{IJ} M^J,
\eeq
with $A_{IJ}$ the intersection matrix on $dP_n$.

Another remarkable fact of this supergravity dual is that it reproduces the field theory 
computation of $n$ combinations of the $n+3$ gauge coupling beta functions
(corresponding to $n$ marginal couplings in the conformal case)  
\cite{Franco:2004jz}. They 
are encoded in the evolution of the NSNS 2-form in the radial direction due to the non-trivial 
NSNS 3-form flux.

The construction of these throats is important since it illustrates that 
cascading RG flows appear often in 
quiver gauge theories. Moreover, warped throats are interesting both from 
the viewpoint of phenomenological 
applications (e.g. \cite{Giddings:2001yu,Kachru:2003sx,Cascales:2003wn}) and of 
counting 
flux vacua, due to their `attractor' behavior \cite{Denef:2004ze}. Our 
purpose in this paper is to clarify the infrared structure of these (and 
similar) classes of models, a key understanding required for the above 
applications.

These throats contain a naked singularity at their origin, and hence are the analogs of the KT throat 
\cite{Klebanov:2000nc} for the conifold. In later sections we will clarify that the dual gauge 
theory infrared dynamics suggests that in many situations a suitable deformation of the geometry 
eliminates the singularity, and yields a smooth supergravity solution, the analog of 
the solution in \cite{Klebanov:2000hb} for the conifold.

The above construction of throats was originally elucidated for the case of del Pezzo surfaces. 
Nevertheless, its range of applicability is much broader and it is indeed suitable for any other
complex cone over a 4-dimensional surface $Y^4$ with a K\"{a}hler-Einstein metric. In the 
general case, the $\phi_I$, with $I=1 \ldots h^{1,1}(Y^4)-1$, correspond to a basis of harmonic $(1,1)$ 
forms chosen to be orthogonal to the K\"{a}hler form on $Y^4$, and $A_{IJ}$ is a general matrix encoding 
the cup product among them (alternatively, the geometric information regarding the intersections 
between the 2-cycles in $Y^4$ which are Poincar\'e dual to the $\phi_I$'s). Indeed, the warped conifold 
belongs to the above class of solution, by considering the case of a single 2-form orthogonal to 
$\omega$, and with the matrix $A_{IJ}$ reduced to a single entry. 

Finally, we would like to mention that there exist more general 
situations, where the conical Calabi-Yau singularity corresponds to a 
real cone over a Sasaki-Einstein 5-dimensional horizon $X_5$ as before, 
but $X_5$ cannot be constructed as a $U(1)$ fibration over a 4-dimensional
K\"{a}hler-Einstein base. Simple examples of this class are provided by 
the complex cones over $dP_1$ and $dP_2$, where the $U(1)$ fibration over 
the del Pezzo surface is irregular. This fact maps, on the gauge theory side, to
irrational R-charges. Moreover, recently, an infinite family 
of cones over 5d Sasaki-Einstein manifolds, denoted $Y^{p,q}$, 
with explicit metrics has been constructed 
\cite{Gauntlett:2004zh,Gauntlett:2004yd,Gauntlett:2004hh,Gauntlett:2004hs,Martelli:2004wu}. 
Also, the dual quiver gauge theories have been found in 
\cite{Benvenuti:2004dy}. Duality cascades for the case of $Y^{2,1}$, 
corresponding to the 5d horizon of a complex cone over $dP_1$, were 
constructed in \cite{Franco:2004jz}, and duality cascades for the entire 
$Y^{p,q}$ family along with their supergravity duals have been 
recently carried out in \cite{Ejaz:2004tr}. An interesting difference with 
respect to the above throats is an additional dependence of the warp 
factor on a coordinate of the 5d horizon $X_5$, rather than just on the 
radial direction.

Finally, notice that, using our 
arguments in coming sections, one can show that the $Y^{p,q}$ cascades do not in 
general admit a geometric deformation to resolve their singularities.
The only cases where this is possible correspond to cones over $Y^{p,0}$, 
which are in fact $\IZ_p$ quotients of the conifold. They thus fall within 
our analysis, and we describe the field theory version of their smoothing 
in Appendix \ref{yp0}.

\subsection{The deformed geometries}
\label{deformedgeom}

The above throats contain a naked singularity, suggesting that they miss 
the non-perturbative infrared dynamics of the dual gauge field theory. 
Hence they are the analogs of the 
singular solution in \cite{Klebanov:2000nc}. From the discussion of the 
conifold it is expected that, at least in some cases, when the infrared 
gauge theory dynamics is included, the dual supergravity solution  
corresponds to a deformed background related to the original one by an 
extremal transition. This transition replaces 2- and 4-cycles by 3-cycles. A 
general question is therefore to analyze the existence of extremal  
transitions on local Calabi-Yau geometries, where shrinking 4-cycles are 
replaced by finite size 3-cycles.

In this section we address this geometric question from several viewpoints. 
For concreteness we center the discussion on the geometries given by 
complex cones over del Pezzo surfaces, although 
results generalize to other situations, as will be clear in our examples.

The general question is what are the possible deformations of the complex cones over del Pezzo 
surfaces. Besides its relevance to the above discussion, this question has another interesting 
realization. Geometries with collapsing del Pezzo surfaces lead, when used as M-theory 
backgrounds, to five-dimensional field theories with $E_n$ global symmetries. The Coulomb branch 
is parametrized by the sizes of the 2-cycles, while the Higgs branch 
corresponds to extremal transitions, i.e. complex deformations of the 
geometry arising at the origin of the Coulomb branch, where the 4-cycle shrinks to zero size. The classification 
of such Higgs branches was described in \cite{Seiberg:1996bd}, and shown
\cite{Morrison:1996xf} to fully agree with the geometric description.\footnote{A concise description of these Higgs branches is provided by 
the instanton moduli space 
of the corresponding $E_n$ gauge theory. The relation is manifest by realizing the 
five-dimensional field theory in the worldvolume of D4-branes probing configurations of 
D8-branes/O8-planes at strong coupling, so that the global symmetry is enhanced to $E_n$ 
\cite{Seiberg:1996bd}. The Higgs branch corresponds to dissolving the D4-brane as an instanton of the $E_n$ gauge theory.}

In many examples, one may use the realization of the five-dimensional 
field theories in terms of $(p,q)$ webs of Type IIB fivebranes  
\cite{Aharony:1997ju, Aharony:1997bh}, in order to visualize the 
corresponding Higgs branches. This corresponds to the situations 
where the geometries are toric, and the $(p,q)$ web corresponds to 
the reciprocal of the collection of points in the $\IZ^2$ integer lattice
defining the toric diagram \cite{Aharony:1997bh,Leung:1997tw}.
In general, for toric geometries with a corresponding web, deformations exist if there are subsets 
of external legs which can form sub-webs in equilibrium. The deformation 
is described as the separation of such sub-webs. A more precise description of this in toric geometry language
is illustrated in some examples in Appendix \ref{toric}.

The $(p,q)$ web representation of the deformation for the conifold is described in Figure \ref{conitrans}, where the sub-webs 
correspond to straight lines. The 3-cycle in the deformed conifold corresponds to a segment stretched between 
the two sub-webs. For example in 5 dimensional gauge theories a D3-brane stretched between the two $(p,q)$ sub-webs 
is a BPS brane on the Higgs branch, which maps to a brane wrapped on the 3-cycle in the geometry.

\begin{figure}[ht]
  \epsfxsize = 8cm
  \centerline{\epsfbox{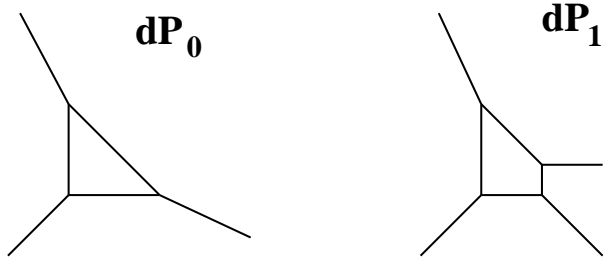}}
  \caption{Web diagrams for the complex cones over $dP_0$ and $dP_1$. In both cases,
it is impossible to split them into more than one sub-webs
in equilibrium, implying there exist no complex deformations for these geometries.}
  \label{dp01}
\end{figure}

Using the toric diagrams for the cones over del Pezzo surfaces one can recover the results in 
\cite{Morrison:1996xf}. Namely, for $dP_0$ and $dP_1$ there is no  
deformation branch, as is manifest from their toric pictures, Figure \ref{dp01}.

\begin{figure}[ht]
  \epsfxsize = 11cm
  \centerline{\epsfbox{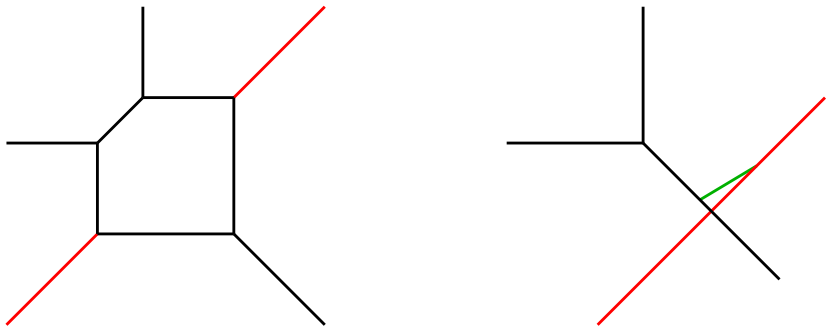}}
  \caption{The web diagram for the complex cone over $dP_2$ and its 
complex deformation.}
  \label{dp2}
\end{figure}

\begin{figure}[ht]
  \epsfxsize = 14cm
  \centerline{\epsfbox{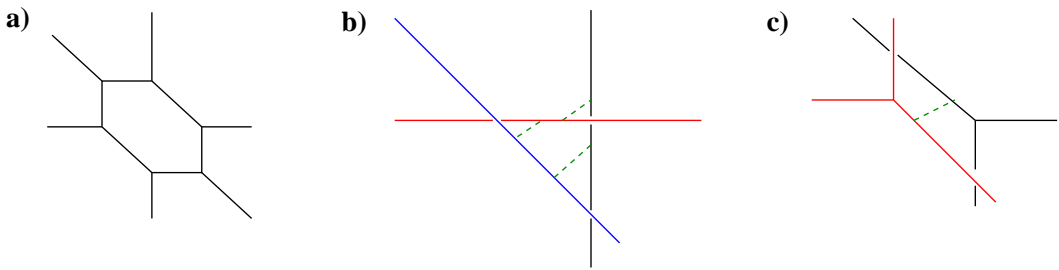}}
  \caption{Web diagram for the complex cone over $dP_3$ and its two 
branches of complex deformation. Figure b) shows a two-dimensional 
deformation branch, parametrized by the sizes of two independent 3-spheres 
corresponding to the dashed segments (the three segments are related by a 
homology relation, hence only two are independent). Figure c) shows a 
one-dimensional deformation branch.}
  \label{dp3}
\end{figure}

On the other hand, $dP_2$ has a deformation, shown in Figure \ref{dp2}, which completely smoothes out the geometry. 
For $dP_3$ there are two deformation branches, one of them two-dimensional and the other one-dimensional, see Figure \ref{dp3}.
Notice that the two-dimensional deformation branch may be regarded as a one-dimensional deformation to the conifold, subsequently
followed by a one-dimensional deformation to a smooth space. This is more manifest in the regime of widely different sizes for the
two independent 3-cycles.

\begin{figure}[ht]
  \epsfxsize = 10cm
  \centerline{\epsfbox{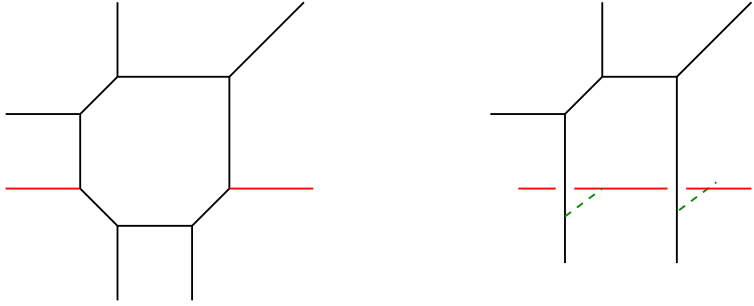}}
  \caption{Web diagram for a cone over a non-generic blow-up of $\IP^2$ 
at four points and its deformation. This geometry is toric and is closely 
related to $dP_4$. The two dashed segments correspond to two homologically 
equivalent 3-spheres. The left-over diagram describes a suspended pinch 
point singularity, which admits a further deformation not shown in the 
picture.}
  \label{web_PdP4}
\end{figure}

For higher del Pezzo surfaces, the generic geometry is not toric. However, 
there are closely related blow-ups of $\IP_2$ at non-generic points, which
do admit a toric description. These non-generic geometries lead to the same quivers
than the del Pezzos, but with different superpotentials. For non-toric del Pezzos, some 
deformations are manifest in the toric representation, see  
\fref{web_PdP4} for an example. Notice however that the dimensions of these deformation 
branches is in general lower than that for generic geometries, thus 
showing that some deformations of the higher del Pezzos are non-toric.

\begin{figure}[ht]
  \epsfxsize = 10cm
  \centerline{\epsfbox{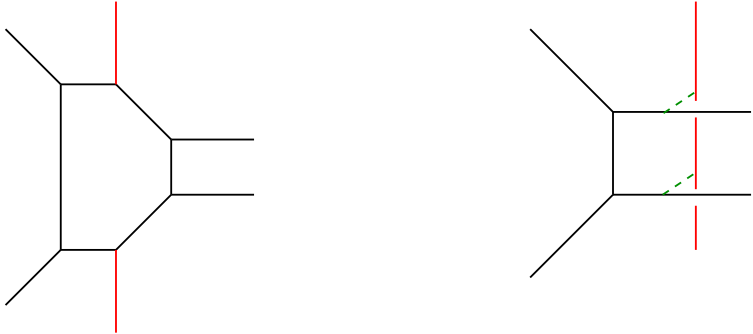}}
  \caption{Web diagram for the cone over a non-generic $dP_3$ and its
deformation to the orbifold $\IC^3/\IZ_2$.}
  \label{web_deformed.eps}
\end{figure}

In a similar spirit, we may consider other toric geometries closely 
related to toric del Pezzos, but corresponding to a non-generic location 
of the blow-ups.\footnote{Here the distinction between toric and quiver 
webs is relevant \cite{Feng:2004uq}. In these
cases, the web diagram corresponds to the quiver web, and encodes the quiver data of a less symmetric phase of
the gauge theory. On the other hand, the geometry is still described by some toric data, corresponding to a 
toric web, different in general from the quiver web. See \cite{Feng:2004uq} for a detailed discussion.}
They are given by web diagrams associated to the so-called less 
symmetric quiver gauge theories. For such geometries, deformations to 
smooth geometries exist, 
although the generic deformation may not be available. One example of a deformation on a 
non-generic version of $dP_3$ is shown in Figure \ref{web_deformed.eps}.

Finally, we emphasize that the above techniques can be used to study the 
deformations of other geometries, even involving more than one collapsing 
4-cycles. Concrete examples, like the deformations
of the cone over $F_0$, the suspended pinch point singularity
and the $Y^{p,q}$ geometries will appear in subsequent sections. It is 
interesting to point out that all possible complex deformations for toric 
varieties may not be described using the above web deformations. Nevertheless, 
all our examples will be of this kind. We leave the interesting question of other 
possible situations for future research.

\subsection{A topological consideration}
\label{topocons}

The throats on cones over $dP_n$ constructed in \cite{Franco:2004jz} (and generalizations) have 
in principle $n$ independent discrete parameters, the $M^I$, associated to the 
integer fluxes sourced by the $n$ independent fractional branes one can in 
principle introduce in the quiver gauge theory.

However, in this section we describe a topological argument which shows 
that in order for a throat to 
have a smooth deformation at its bottom, corresponding to a geometric deformation as 
discussed above, the fractional brane assignments cannot be fully arbitrary.
Equivalently, it is possible to use topological information about the allowed 
deformations to derive the set of fractional branes triggering the corresponding 
strong infrared dynamics. 

The argument is as follows. The fractional brane numbers can be measured 
in the throat solution by computing the flux of the RR 3-form $F_3$ 
through a 3-cycle in the 5d base of the cone (constructed as an $S^1$  
fibration over a 2-cycle in the del Pezzo surface \footnote{Although we focus
on the case of del Pezzo surfaces, the discussion applies to other real four dimensional 
K\"ahler-Einstein surfaces.}). There are $n$ such 
3-cycles. On the other hand, in the smooth deformed geometries, one in 
general finds a smaller number $k$ of 3-cycles. This implies that 
$n-k$ 3-cycles in the asymptotic region are homologically trivial. 
Consequently, only $k$ independent choices of fractional brane numbers 
remain.

For each deformed geometry, the set of corresponding fractional branes, 
i.e. those associated to the homologically non-trivial 3-cycles, is 
determined as follows. Consider a given complex deformation, corresponding 
to the separation of sub-webs. Recall now the relation between external 
legs in web diagram and nodes in the quiver \cite{Hanany:2001py,Franco:2002ae}.
The fractional branes associated to the deformation are those 
controlling the rank of the nodes corresponding to the legs in the 
removed sub-web. We will see some examples of this in later sections. 

Notice that this does not mean there are no throat solutions for more general fractional 
brane assignments, but rather that they cannot be completed in terms of a purely geometric 
background. The most plausible proposal is that the general case corresponds to a smooth deformed 
geometry (accounting for the fractional branes associated to non-trivial 3-cycles) 
with additional explicit fractional brane sources, 
leading to a non-trivial $dF_3$, which induces a non-vanishing $F_3$ flux 
at large 
radial distances for the homologically trivial 3-cycles.


\subsection{Deformations from the gauge theory}
\label{generaldeform}

In the previous sections, we have introduced the simple example of the conifold and discussed how the 
original naked singularity in the supergravity dual is cured when the strong coupling dynamics of the gauge 
theory is taken into account. We then reviewed how more general extremal transitions
are described using toric geometry in the form of $(p,q)$ webs.

We now describe the derivation of the geometric deformation from the 
viewpoint of the infrared dynamics of the dual gauge theory, for a 
general quiver theory. As in the conifold 
case above, the deformation can be derived as the deformed 
moduli space of probes, arising from the quantum modification of the moduli space of the gauge 
theory. Although the basic idea follows discussions in \cite{Klebanov:2000hb}, its implementation 
in our more involved geometries leads to richer structures.

The geometries we study have several collapsing 2-cycles on which we can 
wrap D5-branes, giving rise to different types of fractional branes. In order for the 
supergravity solutions described in Section \ref{sugrathroats} to be valid 
we will assume that the number of fractional branes of each type $M^I \ll N$. There is 
no constraint on the relative sizes of the $M^I$'s. However, in order to simplify our discussion, 
we can consider the situation in which 

\beq
M^1 \ll M^2 \ll \ldots \ll \ldots \ll N
\eeq

Then it is natural to foresee a hierarchy of scales of strong gauge dynamics

\beq
\Lambda_1 \ll \Lambda_2 \ll \ldots \ll \Lambda_3
\eeq
where the $\Lambda_I$'s are dynamical scales that arise when $N(\Lambda_I)$ is comparable to $M_I$ 

\beq
\Lambda_I \mbox{ such that } N(\Lambda_I) \sim M^I
\eeq

We have simplified the field theory analysis by assuming the scales are well separated, although we 
expect that descriptions of other situations exist in both the smooth supergravity solution and 
the gauge theory language.

The basic structure of strong infrared dynamics is the following. Given a quiver gauge theory 
with fractional branes, the theory cascades down until the number of D3-branes $N$ becomes 
similar to one of the fractional brane numbers, say $M^{I_0}$, at a scale $\Lambda_{I_0}$. For 
simplicity, and due to our assumption of separation of scales, we may ignore the remaining 
$M^I$'s and take them to vanish. In order to simplify notation, we call $M^{I_0}=M$. Then the last 
step of the cascade can be probed by introducing $M$ additional D3-branes and studying the resulting 
moduli space. In this situation the
gauge group takes the form
\beq
SU(2M)^m \times SU(M)^n
\eeq

In several of our examples below, the number of gauge factors with rank $2M$ is two 
\footnote{In the $(p,q)$ web description of the deformations presented in Section \ref{deformedgeom}, 
this arises naturally when one of the sub-webs that are separated is simply an infinite straight line.}, $m=2$, but the 
discussion may be carried out in 
general. Also, in the explicit models the number of flavors for the $SU(2M)$ gauge factors is 
$2M$, hence equals the number of colors. 

The non-perturbative dynamics may be determined by assuming, to begin with, that the $SU(M)$ 
gauge factors are weakly coupled and act as spectators, corresponding to global flavor 
symmetries. For simplicity we continue the discussion assuming also that there no arrows among 
$SU(2M)$ nodes, i.e. no $(2M,\overline{2M})$ matter. Under these circumstances, the strong dynamics 
corresponds to a set of decoupled $SU(2M)$ gauge theories with equal number of colors and 
flavors, which thus develop a non-perturbative {\em quantum modification of the moduli space}. 
This is best understood in terms of gauge invariant mesonic and baryonic variables. For 
each such gauge factor, the mesons are

\beq
\mathcal{M}_{ru}=A_r B_u 
\eeq
with $r,u=1,2$, where

\beq
\begin{array}{rlcrl}
A_r:         & (2M,\ov{M}_r) & \ \ \ \ \ \ &         B_u: & (M_u,\ov{2M}) 
\end{array}
\eeq
and the baryons have the abbreviated form 

\beq
\begin{array}{lcl}
\mathcal{B}=[A]^{2M} & \ \ \ \ \ \  & \tilde{\mathcal{B}}=[B]^{2M} 
\end{array}
\eeq
where antisymmetrization of gauge indices is understood. It is important to keep in mind that these operators
are not gauge invariant when the entire gauge group (and not just the factors undergoing deformation) 
is taken into account. This will be important when we study what happens after they develop non-zero
vevs. The quantum modified moduli space is 
described by

\beq
\begin{array}{c}
\det\mathcal{M}-\mathcal{B}\tilde{\mathcal{B}}=\Lambda^{4M} 
\end{array}
\label{qmodif}
\eeq
The resulting infrared gauge dynamics is described by a quiver gauge theory with the $SU(2M)$ 
nodes removed, the corresponding flavors replaced by mesonic and baryonic degrees of freedom (both 
in the quiver diagram and in the superpotential), and with the quantum modified constraints
enforced as superpotential interactions by means of singlet chiral field Lagrange 
multipliers $X$, of the form\footnote{For simplicity, we show only one Lagrange multiplier and additional 
superpotential term. It us understood that there is one such contribution for each strongly coupled 
gauge group factor.} 

\beq
W=W_0+ X (\det\mathcal{M}-\mathcal{B}\tilde{\mathcal{B}}-\Lambda^{4M})
\eeq

The quantum constraints force some of the meson/baryon degrees of freedom 
to acquire non-zero vevs. The dynamics of the probes is recovered along 
the mesonic branch, which corresponds to setting the baryons to zero and 
$X=\pm\Lambda^{4-4M}$, and saturating the constraint with meson vevs (see 
footnote 
\ref{baryon}).
This triggers symmetry breaking of some of the $SU(M)$ 
factors to diagonal combinations, and makes some of the fields massive due to superpotential 
couplings. The resulting theory contains a set of meson fields with quantum deformed moduli space, 
describing the probes in the deformed geometry. In addition, there are additional gauge factors 
and chiral multiplets describing the geometry left-over after the complex structure deformation 
of the original one. In later sections we will present several examples, in which the matching 
between the gauge theory description of the quantum deformations and the geometric complex structure 
deformations is complete. This is a very satisfactory result.

There is a subtlety in fixing the sign of the vev for $X$. The simplest 
way of determining the correct one is to impose that, 
restricting to the Abelian case, the theory has a superpotential allowing for a toric description of its moduli space.
Concretely, that each bi-fundamental field appears with opposite signs in the two terms containing it. This recipe
can be recovered from a more careful treatment of the equation of motion determining $X$ from the initial superpotential,
as discussed in a concrete example in Appendix \ref{mesonic}.

After the condensation, the left-over quiver theory may correspond to a singular geometry with 
fractional branes, and thus will have subsequent duality cascades and condensations. The resulting RG flow takes in this case the form of a cascade with multiple dynamical scales at which the underlying geometry undergoes deformation. Explicit examples are discussed in coming sections.

\section{Some warmup examples}
\label{warmup}

In this section we would like to describe some simple examples of infrared resolutions, in situations 
with one-scale cascades.

\subsection{The cone over $F_0$}

\begin{figure}[ht]
  \epsfxsize = 8cm
  \centerline{\epsfbox{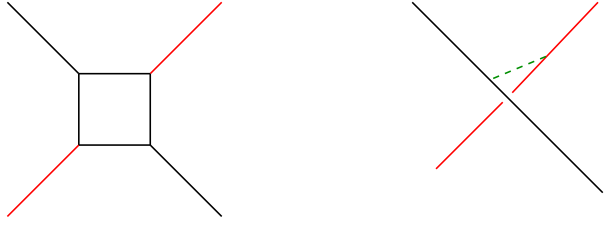}}
  \caption{Web diagram for the complex cone over $F_0$, and its complex deformation to a smooth space.}
  \label{f0}
\end{figure}

\begin{figure}[ht]
  \epsfxsize = 9.5cm
  \centerline{\epsfbox{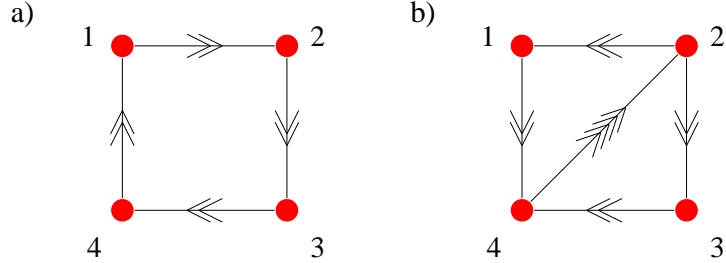}}
  \caption{Figure a) shows the quiver diagram for the theory of D3-branes at a complex cone over $F_0$. Figure b) shows a dual 
phase of the theory, involved in the duality cascade.}
  \label{quiverf0}
\end{figure}

Let us consider the case of the cone over $F_0$. The web diagram for this geometry is shown in
Figure \ref{f0}a, and the corresponding quiver is in Figure \ref{quiverf0}a.

The fractional brane corresponds to the rank vector $(0,1,0,1)$.
The superpotential for the theory is
\beqa
W & = & X_{12} X_{23} Y_{34} Y_{41} \, -\, X_{12} Y_{23} Y_{34} X_{41} \, -
 Y_{12}X_{23} X_{34} Y_{41} \, + \, Y_{12} Y_{23} X_{34} X_{41}
\eeqa
in self-explanatory notation. This theory has an $SU(2) \times SU(2)$ global 
symmetry, which geometrically arises as the product of the $SU(2)$ isometries
of the two $\IP^1$'s in $F_0=\IP^1 \times \IP^1$.
It corresponds to a $\IZ_2$ orbifold of the conifold $xy-zw=0$ by the 
action $x,y,z,w\to -x,-y,-z,-w$, 
as first determined in \cite{Morrison:1998cs}. This is also
manifest in the dual toric diagrams, where the cone over $F_0$ differs from the conifold by the addition of
an interior point (namely, by the refinement of the toric lattice).

This theory has a cascade, which was exhaustively discussed in \cite{Franco:2003ja,Franco:2004jz}, to which we refer the reader
for details. Introducing $N$ D3-branes and $M$ fractional branes, namely starting quiver \ref{quiverf0}a with the rank vector
\beqa
N\, (1,1,1,1)\, +\, M\, (0,1,0,1)
\eeqa
the theory alternates between the two quivers in Figure \ref{quiverf0}a, b. 
Given the $\IZ_2$ symmetry of the quiver and of the deformed geometry, it 
is natural to consider the situation where the UV 
gauge couplings of opposite nodes are equal. In this case, the duality cycle is obtained by a (simultaneous) 
dualization of the nodes 1 and 3, followed by a (simultaneous) dualization of 2 and 4, after which 1 and 3
are subsequently dualized, etc. Under these conditions, quiver \ref{quiverf0}b appears just as an intermediate 
step between simultaneous dualizations. 
In each duality cycle, the number of D3-branes decreases by $2M$ units.
It is interesting to note that, regarding the geometry as a quotient of the conifold, nodes 1 and 3 descend from a single node in the quiver 
of the conifold, while 2 and 4 descend from the other. In this respect, the duality cascade in the orbifold theory, in the situation of 
symmetric gauge couplings for opposite nodes in the quiver,  can be regarded as directly descending from the duality cascade in the 
conifold theory.

The infrared end of the cascade is therefore expected to be similar to
that of the conifold. In fact, this is exactly what is obtained e.g. for $N$ a multiple of $M$. The
gauge theory associated to the rank vector $(0,M,0,M)$, leads to two 
decoupled ${\cal N}=1$ SYM-like theories. 
In more detail, the infrared behavior
may be explored by introducing $M$ additional D3-brane probes, namely by
studying the gauge theory with rank vector $(M,2M,M,2M)$. In the 
infrared
the gauge factors 1 and 3 are weakly coupled, and can be considered
spectators. The gauge factors 2 and 4 have
$N_f=N_c$ and develop a quantum deformation of their moduli space.
Following the general discussion in Section \ref{generaldeform}, we introduce the mesons
\beqa
{\cal M} & = & 
\left[ \begin{array}{cc} M_{11} \, & \, M_{12} \cr
M_{21}\, & \, M_{22} \end{array} \right] \, =\,
\left[ \begin{array}{cc} X_{12}X_{23} \, & \, X_{12}Y_{23} \cr
Y_{12} X_{23} \, & \, Y_{12}Y_{23} \end{array} \right] 
\quad ; \quad
{\cal N}  =  
\left[ \begin{array}{cc} N_{11} \, & \, N_{12} \cr
N_{21} \, & \, N_{22} \end{array} \right]\, =\,
\left[ \begin{array}{cc} X_{34}X_{41} \, & \, X_{34}Y_{41} \cr
Y_{34} X_{41} \, & \, Y_{34}Y_{41} \end{array} \right] \nonumber
\eeqa
and the baryons ${\mathcal{B}}$, ${\tilde {\mathcal B}}$, ${\mathcal C}$, ${\tilde {\mathcal C}}$

The quantum modified superpotential reads
\beqa
W & = & M_{11} N_{22} \, -\,  M_{12}N_{21}\, -\,  M_{21} N_{12}\, +\, 
M_{22} N_{11} \, +\nonumber\\
& + & X_1\, (\det {\mathcal M} - {\mathcal B}{\tilde {\mathcal B}}-\Lambda^{4M})\, 
    +\, X_2\, (\det {\mathcal N} - {\mathcal C}{\tilde {\mathcal C}}-\Lambda^{4M})
\eeqa
where we have introduced a single strong coupling scale due to the equality of the gauge couplings along the flow.

In order to study the mesonic branch, we have
\beqa
& X_1=\Lambda^{4-4M} \quad ; \quad \mathcal{B}=\tilde{\mathcal{B}}=0 \quad 
; \quad 
X_2=\Lambda^{4-4M} \quad ; \quad \mathcal{C}=\tilde{\mathcal{C}}=0 & 
\nonumber \\
& \det\mathcal{M}=\Lambda^{4M}\quad ; \quad  \det\mathcal{N}=\Lambda^{4M}
\eeqa

Now restricting to the Abelian case (see footnote \ref{U1issue}), the 
resulting superpotential is
\beqa
W & = & M_{11} N_{22} \, -\,  M_{12}N_{21}\, -\,  M_{21} N_{12}\, +\, 
M_{22} N_{11}- 
M_{11} M_{22}\, +\, M_{12} M_{21}\,  -\,N_{11}N_{22}\, +\, N_{12} N_{21}
\nonumber
\eeqa
That is, the superpotential becomes entirely quadratic.
The gauge group is broken to the diagonal combination of nodes 1 and 3
by the expectation values of the mesons.
Using the equations of motion, the superpotential vanishes. The only
degrees of freedom are one set of mesons, due to the equations of 
motion, which require ${\cal M}={\cal N}$. In addition, these mesons are 
subject to the quantum constraints, namely $\det {\cal M}=\Lambda^4$.
This describes the dynamics of the probes in the deformation of the cone
over $F_0$ to a smooth space. Indeed, at any point in the mesonic branch 
(in the Abelian case) the gauge group is $U(1)$ and there are three adjoint
(i.e. uncharged) chiral multiplets with vanishing superpotential. This is 
the ${\cal N}=4$ $U(1)$ SYM of D3-branes probing a smooth space.

The analogy of the above discussion with the conifold case is manifest from the
orbifold description. Moreover, from the geometric viewpoint the 
deformation of the cone over $F_0$ corresponds simply to the quotient of 
the deformed conifold $xy-zw=\epsilon$ by the $\IZ_2$ action $x,y,z,w\to 
-x,-y,-z,-w$, under which it is invariant.

\medskip

The complex cone over $F_0$ is one of the first examples in the 
family of real cones over the manifolds $Y^{p,0}$ introduced in 
\cite{Gauntlett:2004zh,Gauntlett:2004yd,Gauntlett:2004hh,Gauntlett:2004hs,Martelli:2004wu},
namely $Y^{2,0}$. The real cones over $Y^{p,0}$ correspond to quotients of
the conifold $xy-zw=0$ by the $\IZ_p$ action generated by
\beqa 
x\to e^{2\pi i/p}x\quad , \quad
y\to e^{-2\pi i/p}y\quad , \quad
z\to e^{2\pi i/p}z\quad , \quad
w\to e^{-2\pi i/p}w
\label{quotient}
\eeqa
(with $Y^{1,0}$ corresponding to $T^{1,1}$, the base of the conifold itself).
This orbifold action is easily understood by looking at the toric diagrams for these varieties.
The toric diagrams look like the diagram of the conifold with an additional refinement of 
the lattice. 

Moreover, using the web diagrams for these varieties it follows that these 
are the only examples of cones over the manifolds $Y^{p,q}$ which admit a 
complex deformation which smoothes the singularity. Namely, only for the case 
of $q=0$ we expect that complex deformations will smooth out the naked 
singularity at the tip of the warped throat solutions in the presence of  
fluxes as in \cite{Ejaz:2004tr}. The discussion of the geometries involved 
and the field theory description of the smoothing is presented in 
Appendix \ref{yp0}.

\subsection{First del Pezzos}

\label{section_first_del_Pezzos}

Let us consider the cones over the first del Pezzo surfaces. As already mentioned, the cone over $dP_0$ does
not admit any fractional branes, and therefore cannot be taken away from the conformal regime. 
 
\begin{figure}[ht]
  \epsfxsize = 3cm
  \centerline{\epsfbox{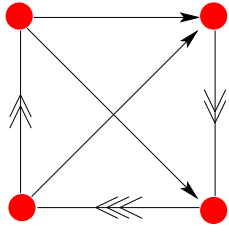}}
  \caption{Quiver diagram for D3-branes at the cone over $dP_1$.}
  \label{quiver_dP1}
\end{figure}

The quiver diagram for a cone over $dP_1$ is presented in \fref{quiver_dP1}. The corresponding superpotential is

\beq
W=\epsilon_{\alpha \beta} X^{\alpha}_{34} X^{\beta}_{41} X_{13}-\epsilon_{\alpha \beta} X^{\alpha}_{34} X_{42} X^{\beta}_{23}+
\epsilon_{\alpha \beta} X^{12} X^3_{34} X^{\alpha}_{41} X^{\beta}_{23}
\label{W_dP1}
\eeq

This theory admits one kind of fractional branes, given by the rank vector 
$(0,3,1,2)$. The addition of these fractional branes leads to an RG cascade 
which was first studied in \cite{Franco:2004jz}. The superpotential \eref{W_dP1} 
preserves an $SU(2) \times U(1)$ global symmetry. The R-charges can then be 
determined using the a-maximization 
principle, and turn out to be irrational numbers \cite{Martelli:2004wu}. Some explicit computations can be found in \cite{Bertolini:2004xf}. This 
is the simplest example of a singularity whose dual gauge theory has irrational R-charges. Thus, it is
very interesting to understand the associated cascades in detail and we now proceed to do so.

\begin{figure}[ht]
  \epsfxsize = 12cm
  \centerline{\epsfbox{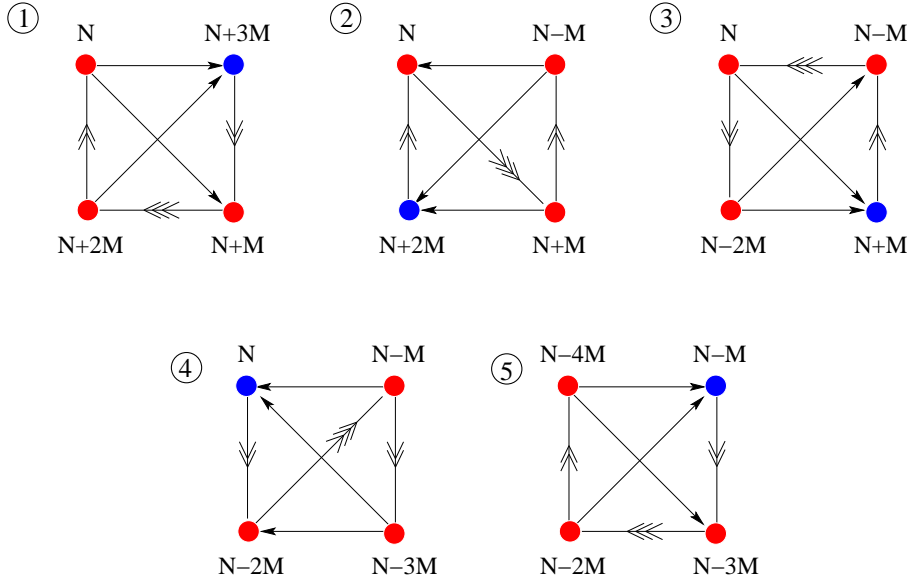}}
  \caption{Quivers in a duality cycle in the duality cascade of $dP_1$. We have indicated in blue the dualized node at each step.}
  \label{cascade_quivers_dP1}
\end{figure}

The resulting RG flow is logarithmic and periodic. For an appropriate choice of initial couplings, the 
sequence of dualized nodes in a period is 2, 4, 3, 1, after which $N \rightarrow N-4M$ and $M$. The 
quivers for several steps in the cascade are shown in \fref{cascade_quivers_dP1}. 

The beta functions at each step are

{\footnotesize
\beq
\begin{array}{l|cccc|cccc}
            &  \ \ \ \ N_1 \ \ \ \ &  \ \ \ \ N_2 \ \ \ \ &  \ \ \ \ N_3 \ \ \ \ &  \ \ \ \ N_4 \ \ \ \ &
               \ \ \ \beta_1/M \ \ \ & \ \ \ \beta_2/M \ \ \ & \ \ \ \beta_3/M \ \ \ & \ \ \ \beta_4/M \ \ \ \\
\hline 
\hline 
{\bf 1} \ \ &    N &  N+3M &   N+M &  N+2M & -10+\sqrt{13}     & \bf{10-\sqrt{13}} & 22-7\sqrt{13}     & -22+7\sqrt{13}    \\
{\bf 2} \ \ &    N &   N-M &   N+M &  N+2M & 22-7\sqrt{13}     & -10+\sqrt{13}     & -22+7\sqrt{13}    & \bf{10-\sqrt{13}} \\
{\bf 3} \ \ &    N &   N-M &   N+M &  N-2M & -22+7\sqrt{13}    & 22-7\sqrt{13}     & \bf{10-\sqrt{13}} & -10+\sqrt{13}     \\
{\bf 4} \ \ &    N &   N-M &  N-3M &  N-2M & \bf{10-\sqrt{13}} & -22+7\sqrt{13}    & -10+\sqrt{13}     & 22-7\sqrt{13}     \\
{\bf 5} \ \ &    N &  N-4M &  N-3M &  N-2M & -10+\sqrt{13}     & \bf{10-\sqrt{13}} & 22-7\sqrt{13}     & -22+7\sqrt{13}
\end{array}
\label{cascade_dP1_betas}
\eeq
}
where we have indicated the beta functions of the dualized nodes with a bold font.

\begin{figure}[ht]
  \epsfxsize = 7cm
  \centerline{\epsfbox{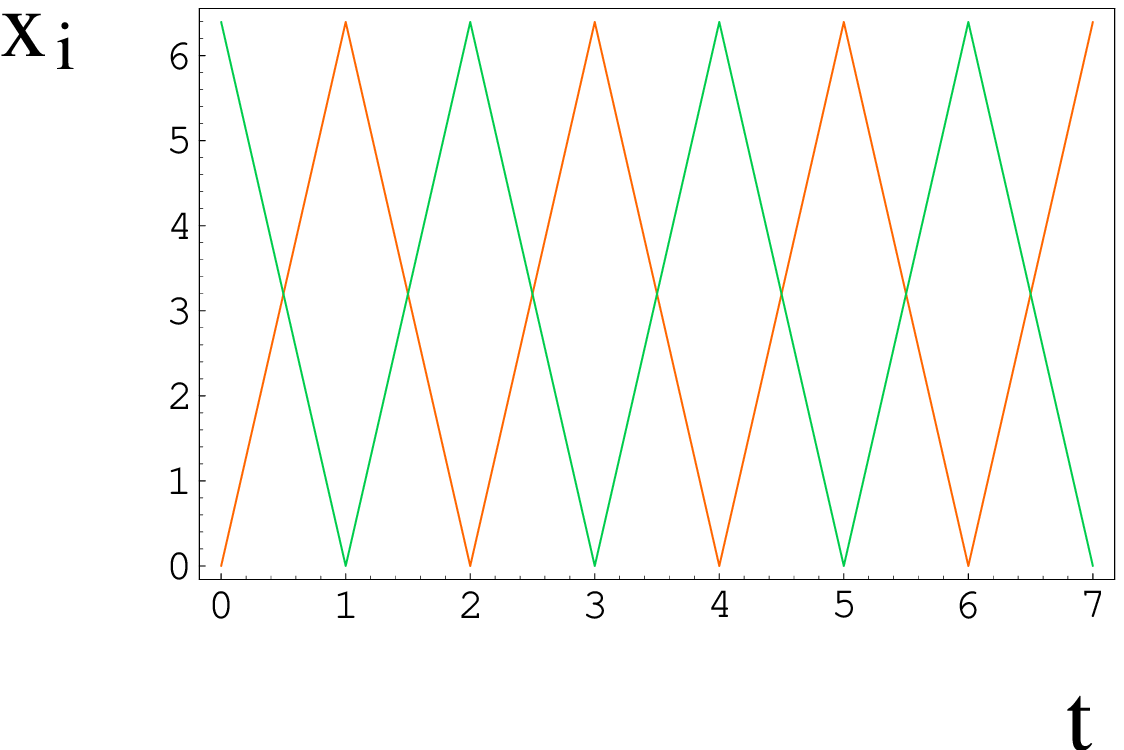}}
  \caption{Evolution of the inverse squared couplings $x_i=8 \pi^2/g_i^2$ as a function of $t=\log \mu$ for the $dP_1$ cascade under consideration.
UV couplings have been chosen respecting the quiver symmetries and such that the sequence given by \fref{cascade_quivers_dP1} and 
\eref{cascade_dP1_betas} is followed. We indicate $x_1$ and $x_2$ in green, and $x_3$ and $x_4$ in orange.}
  \label{couplings_cascade_dP1}
\end{figure}

In addition, the supergravity dual of this flow corresponds to the $Y^{2,1}$ flow, which is a member of the class of warped 
throat solutions recently constructed in 
\cite{Ejaz:2004tr}. However, as already mentioned, the geometry does not admit a complex deformation, hence the
naked singularity at the infrared is not removed by this mechanism. This
remains an open question we hope to address in the future.

\medskip

\begin{figure}[ht]
  \epsfxsize = 4cm
  \centerline{\epsfbox{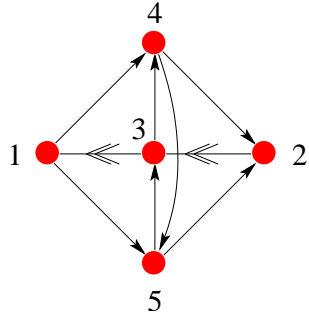}}
  \caption{Quiver diagram for D3-branes at the cone over $dP_2$.}
  \label{quiverdp2}
\end{figure}

The first non-trivial example of complex deformation is provided by the cone over $dP_2$.
The web diagram is shown in Figure \ref{dp2}a, and the corresponding quiver diagram is in Figure \ref{quiverdp2}.

The superpotential for this theory is given by
\beqa
W & = & X_{34}X_{45}X_{53}-(X_{53}Y_{31}X_{15}+X_{34}X_{42}Y_{23}) 
\nonumber\\
  & + & (Y_{23}X_{31}X_{15}X_{52}+X_{42}X_{23}Y_{31}X_{14})-X_{23}X_{31}X_{14}X_{45}X_{52}
\label{W_dP2_1}
\eeqa

The two independent fractional branes can be taken to correspond to the rank vectors $(1,1,0,0,0)$ and $(0,1,0,1,-1)$.
The existence of an RG cascade in this theory, although expected, has not 
been established in the literature, neither from the field 
theory nor the supergravity viewpoint. 

Using our arguments in section \ref{topocons}, it is possible to see that the cascade 
ending in the deformation shown in figure \ref{dp2}a corresponds to the first type 
of fractional branes. We thus proceed to study it, taking initial ranks of the 
form
\beq
\vec{N}=N(1,1,1,1,1)+M(1,1,0,0,0)
\eeq
We will consider UV couplings respecting the $\IZ_2$ symmetry that the quiver has in the absence of fractional
branes, $x_1=x_2$ and $x_4=x_5$. 

\begin{figure}[ht]
  \epsfxsize = 14cm
  \centerline{\epsfbox{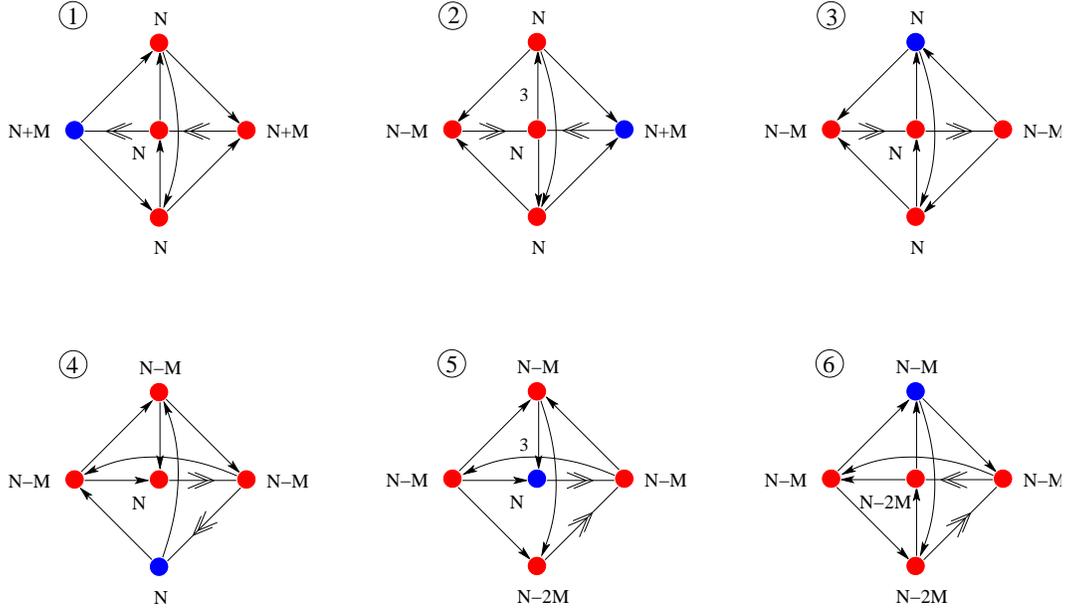}}
  \caption{Some quivers in a duality cycle in the duality cascade of $dP_2$. We have indicated in blue the dualized node at each step.}
  \label{cascade_quivers_dP2}
\end{figure}

The sequence of gauge group ranks and beta 
functions for the gauge couplings is

{\footnotesize
\beq
\hspace{-1.2cm}
\begin{array}{l|ccccc|ccccc}
            &  \ \ \ \ N_1 \ \ \ \ &  \ \ \ \ N_2 \ \ \ \ &  \ \ \ \ N_3 \ \ \ \ &  \ \ \ \ N_4 \ \ \ \ &   \ \ \ \ N_5 \ \ \ \ &
               \ \ \ \beta_1/M \ \ \ & \ \ \ \beta_2/M \ \ \ & \ \ \ \beta_3/M \ \ \ & \ \ \ \beta_4/M \ \ \ & \ \ \ \beta_5/M \ \ \ \\
\hline 
\hline 
{\bf 1} \ \ &    N+M &  N+M & N &  N & N & \bf{3} & 3 & {3\over 4}(-9+\sqrt{33}) & {3\over 8}(1-\sqrt{33}) & {3\over 8}(1-\sqrt{33})   \\
{\bf 2} \ \ &    N-M &  N+M & N &  N & N & -3 & \bf{3} & 0 & 0 & 0 \\
{\bf 3} \ \ &    N-M &  N-M & N &  N & N & -3 & -3 & {3\over 4}(9-\sqrt{33}) & \bf{{3\over 8}(-1+\sqrt{33})} & {3\over 8}(-1+\sqrt{33})      \\
{\bf 4} \ \ &    N-M &  N-M & N &  N-M & N & {3\over 8}(1-\sqrt{33}) & {3\over 4}(-9+\sqrt{33}) & 3 & {3\over 8}(1-\sqrt{33}) & \bf{3}   \\
{\bf 5} \ \ &    N-M &  N-M & N &  N-M & N-2M & 0 & 0 & \bf{3} & 0 & -3 \\
{\bf 6} \ \ &    N-M &  N-M & N-2M &  N-M & N-2M & {3\over 8}(-1+\sqrt{33}) & {3\over 4}(9-\sqrt{33}) & -3 & \bf{{3\over 8}(-1+\sqrt{33})} & -3
\nonumber
\end{array}
\label{cascade_dP2_betas}
\eeq
}

\begin{figure}[ht]
  \epsfxsize = 7cm
  \centerline{\epsfbox{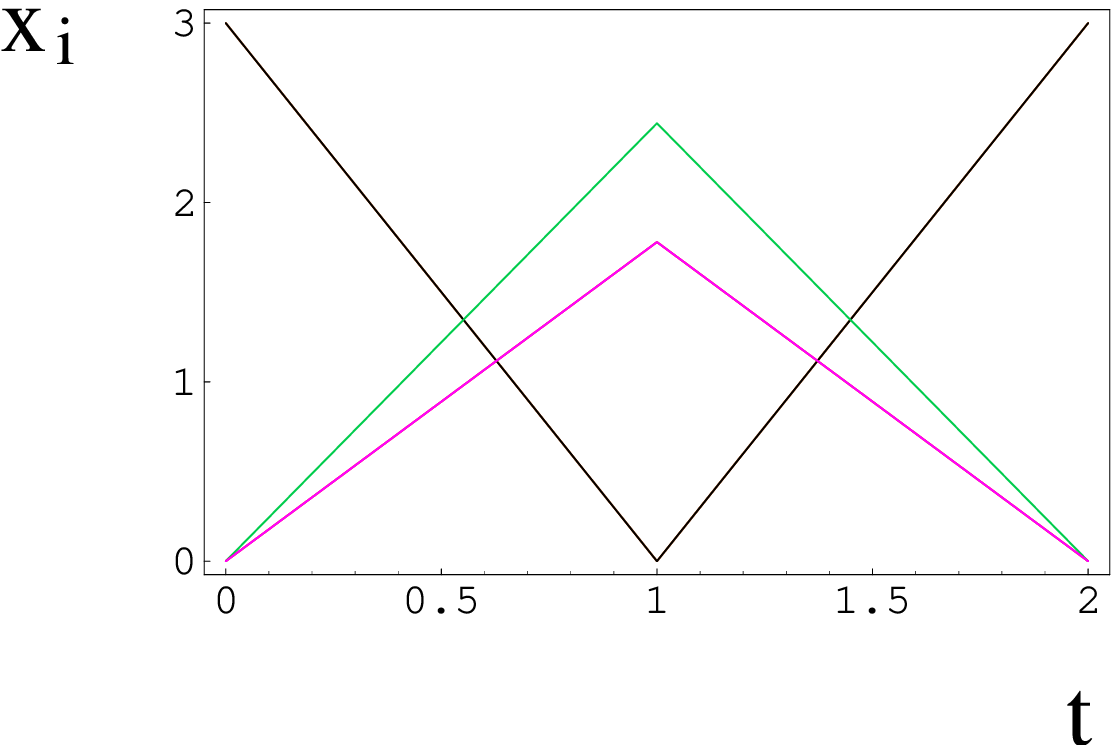}}
  \caption{Evolution of the inverse squared couplings $x_i=8 \pi^2/g_i^2$ as a function of $t=\log \mu$ for some steps in the $dP_2$ 
cascade under consideration. UV couplings have been chosen respecting the quiver symmetries and such that the sequence given by 
\fref{cascade_quivers_dP2} and \eref{cascade_dP2_betas} is followed. We indicate $x_1$ and $x_2$ in black, $x_3$ in green, and 
$x_4$ and $x_5$ in magenta.}
  \label{couplings_cascade_dP2}
\end{figure}

\fref{couplings_cascade_dP2} shows a typical 
evolution of gauge couplings in this case. 
For simplicity, \fref{cascade_quivers_dP2} and \eref{cascade_dP2_betas} 
only show six steps in the duality cascade. At the end of this pattern of 
dualization, one obtains a 
quiver similar to the original one, up to a reduction of the number of D3-branes and a 
rotation of the diagram. Hence continuation of this pattern eventually 
leads to a full duality cycle, and thus a periodic cascade.

Let us now explore the behavior of the theory for small number 
of regular D3-branes, which corresponds to the infrared of the RG cascade. 
For that, we consider $M$ D3-branes probing the theory at the IR end of 
the cascade. Hence, let us consider the gauge theory described by the rank 
vector
\beq
\vec{N}=M(1,1,1,1,1)+M(1,1,0,0,0)
\eeq 
In this situation the nodes 1 and 2 have $N_f=N_c$ and develop a quantum 
deformed moduli space. The meson fields for nodes 1 and 2 are
\beqa
{\cal M}=\left[\begin{array}{cc} M_{34} & M_{35} \\ \tilde{M}_{34} & \tilde{M}_{35} \end{array}\right] =
\left[\begin{array}{cc} X_{31} X_{14} & X_{31} X_{15} \\ Y_{31} X_{14} & Y_{31} X_{15}
\end{array}\right]
\quad ; \quad
{\cal N}=
\left[\begin{array}{cc} N_{43} & N_{53} \\ \tilde{N}_{43} & \tilde{N}_{53} \end{array}\right]=
\left[\begin{array}{cc} X_{42} X_{23} & X_{52} X_{23} \\ X_{42} Y_{23} & X_{52} Y_{23}
\end{array}\right] 
\nonumber
\eeqa

The quantum modified superpotential becomes
\beqa
W & = & X_{34}X_{45}X_{53}-(X_{53}Y_{31}X_{15}+X_{34}X_{42}Y_{23}) 
\nonumber\\
  & + & 
(Y_{23}X_{31}X_{15}X_{52}+X_{42}X_{23}Y_{31}X_{14})-X_{23}X_{31}X_{14}X_{45}X_{52} 
\nonumber\\
  & + & X_1\, (\det {\mathcal M} - {\mathcal B}{\tilde {\mathcal B}}-\Lambda^{4M})\, 
    +\, X_2\, (\det {\mathcal N} - {\mathcal C}{\tilde {\mathcal C}}-\Lambda^{4M})
\label{W_dP2_2}
\eeqa
Along the mesonic branch we have
\beqa
& X_1=\Lambda^{4-4M} \quad ; \quad \mathcal{B}=\tilde{\mathcal{B}}=0 \quad 
; \quad 
X_2=-\Lambda^{4-4M} \quad ; \quad \mathcal{C}=\tilde{\mathcal{C}}=0 & 
\nonumber \\
& \det\mathcal{M}=\Lambda^{4M}\quad ; \quad  \det\mathcal{N}=\Lambda^{4M}
\label{vevs_dP2}
\eeqa

The appropriate signs for the vevs for $X_1$ and $X_2$ can be determined with a reasoning 
identical to the one in Appendix \ref{mesonic}.

The expectation values for the mesons higgs the gauge group to a single diagonal combination 
of the nodes 3, 4 and 5. Restricting to the Abelian case, the superpotential becomes
\beqa
W & = & X_{34}X_{45}X_{53}-N_{53}M_{34}X_{45}-X_{53}\tilde{M}_{35}-X_{34}\tilde{N}_{43} \nonumber \\
  & + & \tilde{N}_{53} M_{35}+N_{43} \tilde{M}_{34}+M_{34} \tilde{M}_{35}-\tilde{M}_{34} M_{35}-N_{43}\tilde{N}_{53}+\tilde{N}_{43} 
N_{53}
\label{W_dP2_3}
\eeqa

Using the equations of motion for e.g. $\tilde{M}_{34}$, $\tilde{M}_{35}$ 
and $\tilde{N}_{43}$, we have
\beqa
M_{34}=X_{53} \quad , \quad M_{35}=N_{43} \quad , \quad X_{33}=N_{53}
\eeqa
Plugging this into \eref{W_dP2_3} we have
\beq
W = X_{34}X_{45}X_{53}-X_{34}X_{53}X_{45}
\label{W_dP2_4}
\eeq
Renaming $X_{34}=X$, $X_{45}=Y$ and $X_{53}=Z$, we obtain the $\mathcal{N}=4$ field content and superpotential
\beq
W=X[Y,Z]
\eeq
which in any event vanishes in the Abelian case, but is crucial in non-Abelian situations. Hence, the moduli space of 
the D3-brane probes is given by the complex deformation of the cone over $dP_2$ to a smooth space, as expected from the geometrical 
analysis.

\subsection{The suspended pinch point}
\label{thespp}

\begin{figure}[ht]
  \epsfxsize = 8.5cm
  \centerline{\epsfbox{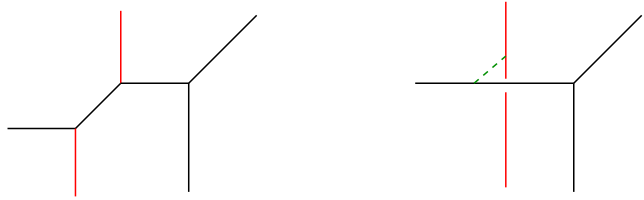}}
  \caption{Web diagram for the SPP and its deformation to a smooth geometry.}
  \label{web_SPP}
\end{figure}

To illustrate that the ideas of cascades and infrared deformations are 
very general, we would like to consider a further example, based on the 
suspended pinch point (SPP) singularity. The web diagram for this 
geometry is shown in Figure \ref{web_SPP}a, while its deformation is in 
Figure \ref{web_SPP}b.

\begin{figure}[ht]
  \epsfxsize = 3.3cm
  \centerline{\epsfbox{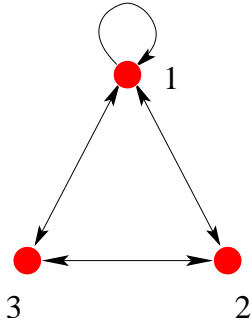}}
  \caption{Quiver diagram for SPP.}
  \label{quiver_SPP}
\end{figure}

The quiver diagram was determined in \cite{Morrison:1998cs,Uranga:1998vf} and is shown in Figure \ref{quiver_SPP}a, and the 
superpotential is
\beq
W=X_{21}X_{12}X_{23}X_{32}-X_{32}X_{23}X_{31}X_{13}+X_{13}X_{31}X_{11}-X_{12}X_{21}X_{11}
\eeq
The ranks of the gauge factors are arbitrary, hence there are two independent fractional branes, which can
be taken to be $(0,1,0)$ and $(0,0,1)$.

Although it has not been described in the literature, the theory has a 
very nice and simple sequence of dualities, which as we show ends in the 
deformed geometry shown in Figure \ref{web_SPP}a. Similarly to what happens in 
the flows considered for $dP_1$ and $dP_2$, this sequence of dualities shares a very 
special feature with the conifold cascade: it is periodic and involves a 
single quiver. Let us consider the starting point given by the ranks
\beq
\vec{N}=N(1,1,1)+M(0,1,0)
\eeq

\begin{figure}[ht]
  \epsfxsize = 11cm
  \centerline{\epsfbox{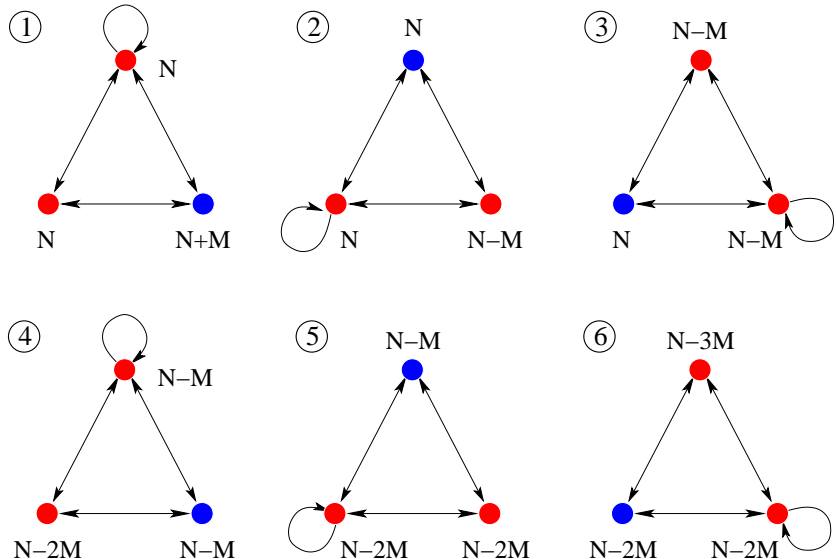}}
  \caption{Sequence of quivers in one period of the SPP cascade. We have indicated in blue the dualized node at each step.}
  \label{quivers_cascade_SPP}
\end{figure}

A period in the duality sequence involves the following set of consecutive dualizations $(2,1,3,2,1,3)$. After six dualizations, the quiver comes back to itself, with $N \rightarrow N-3M$ and $M$ constant. The quiver theories at each step of this sequence are shown in \fref{quivers_cascade_SPP}. As before, we have indicated in blue the node that gets dualized at each step.

Computing the beta functions for the gauge couplings, it appears that the simple sequence in \fref{quivers_cascade_SPP} cannot be realized by a cascade. In particular, it is not possible to prevent a node with an adjoint from going to infinite coupling at some point in the RG flow. It would be interesting to understand whether the simple features of \fref{quivers_cascade_SPP} can be preserved by some more general RG flow.

When the effective number of D3-branes is comparable to $M$, we expect the gauge theory strong dynamics to take over and induce a geometric transition. Indeed, the SPP singularity admits a 
complex deformation, shown in Figure \ref{web_SPP}b. In the following we describe how this arises in the field theory.

In order to study the infrared end of the cascade, we study the gauge theory describing $M$ D3-branes probing it. This corresponds 
to the quiver theory with rank vector
\beq
\vec{N}=M(1,1,1)+M(0,1,0)
\eeq

In this case, we only need to consider mesons and baryons for node 2. The mesons are given by
\beq
{\mathcal{M}} = \left[\begin{array}{cc} M_{13} & M_{11} \\  
M_{33} & M_{31}\end{array} \right]
            = \left[\begin{array}{ccc} X_{12}X_{23} & \ & X_{12}X_{21} \\ X_{32}X_{23} & \ & 
X_{32}X_{21} \end{array} \right]
\eeq

We now introduce the quantum constraint in the superpotential and choose the mesonic branch
\beqa
X=\Lambda^{4-4M}\quad ;\quad {\mathcal B}={\tilde {\mathcal B}}=0
\eeqa
Restricting to the Abelian case, the superpotential reads
\beqa
W&=&- M_{33} X_{31} X_{13} + X_{13}X_{31} X_{11} - M_{11} X_{11} + M_{33}M_{11}
\eeqa
The equation of motion for $M_{11}$ requires $X_{11}=M_{33}$, so we get
\beqa
W&=&- M_{33} X_{31} X_{13} + X_{13}X_{31} M_{33}
\eeqa
The gauge group is $SU(M)$ (due to the breaking by meson vevs 
$M\propto\id$). All three
fields transform in the adjoint representation (a singlet in the Abelian case). The above theory clearly
describes the field content and superpotential of ${\cal N}=4$ SYM, i.e. the theory describing the smooth 
geometry left over after the deformation.

In addition, there remain some additional light fields, namely $M_{11}$, $M_{13}$, $M_{31}$, 
$M_{33}$, subject to the constraint
\beqa
M_{13}M_{31}-M_{33}M_{11}=\Lambda^4
\eeqa

The dynamics is that of probe D3-branes in the geometry corresponding to the deformation of the 
SPP to flat space.
This matches nicely the geometric expectation, from the web diagrams in \fref{web_SPP}, from which
we see that the result of the deformation is a smooth geometry.

\medskip

The relation between the field theory and  the more geometrical description of the deformation can 
be done also using the toric geometry language. Using the construction of the moduli space of the SPP
in terms of toric data (the forward algorithm), e.g. in \cite{Morrison:1998cs,Park:1999ep}, the moduli space 
is given by $xy=zw^2$, with

\beqa
x = X_{13} X_{32} X_{24}, \ \ \ \ y = X_{31} X_{12} X_{23}, \ \ \ \ z = X_{11}, \ \ \ \ w = X_{13} X_{31} 
\eeqa
modulo relations from the superpotential (namely, we also have e.g. $w = 
X_{12} X_{21}$). Using the mesons we have
\beqa
x = X_{13} M_{31}\; ,\; y = X_{31}M_{13}\; ,\; z=X_{11}=M_{33}\; ,\; 
w=X_{13}X_{31}=M_{11}
\eeqa
The monomials satisfy $xy-zw^2=0$ at the classical level, namely
\beqa
X_{13}X_{31} (M_{31}M_{13}-M_{33}M_{11})=0
\eeqa
The quantum deformation of the moduli space of the field theory 
$M_{31}M_{13}-M_{33}M_{11}=\Lambda^4$, thus corresponds to 
\beqa
X_{13}X_{31} (M_{31}M_{13}-M_{33}M_{11})=\epsilon X_{13}X_{31}
\eeqa
which in terms of the monomials can be written as $xy-zw^2=\epsilon w$ 
which is the description of the geometric deformation in Figure 
\ref{web_SPP}b. Thus the description we have provided has a quite direct 
link with the geometric description of the deformation, see Appendix 
\ref{toric}. Similar computations could be carried out in the other 
cases.

\section{The $dP_3$ example}
\label{dpthree}

In this and subsequent sections we present examples where there are several scales of strong gauge dynamics along the
RG flow. They are dual to supergravity solutions with several geometric features along the radial direction. The cleanest examples 
are those involving several deformation scales, which separate throat-like regions with different warp factors, dual to 
cascading flows in the gauge theory. In this section we center on one such example, based on the cone over $dP_3$.

The complex cone over $dP_3$ has two different deformation branches, shown in Figure \ref{dp3}.
Following the discussion in section \ref{topocons}, it is possible to directly determine the sets 
of fractional branes in the gauge theory that are associated to the finite size 3-cycles in the 
supergravity description, and which should therefore trigger the corresponding RG flow and strong 
dynamics. In this section we carry out the gauge theory analysis corresponding to these sets of 
fractional branes and describe in detail the duality cascade and infrared dynamics.

\begin{figure}
\begin{center}
\centering
\epsfysize=3.5cm
\leavevmode
\epsfbox{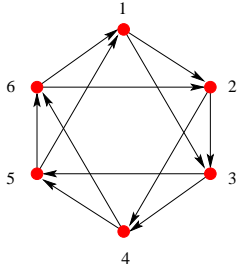}
\end{center}
\caption[]{\small The quiver for D3-branes on the complex cone over
$dP_3$.}
\label{quiverdp3}
\end{figure}

Before doing that, let us review some general features of the gauge 
theory. The $(p,q)$ web diagram is shown in Figure \ref{dp3}, and the 
corresponding quiver gauge theory is shown in figure \ref{quiverdp3}.
The superpotential reads (see e.g. \cite{Feng:2002zw})
\beqa
W & = & X_{12} X_{23} X_{34} X_{45} X_{56} X_{61} \, +\,  X_{13} X_{35} X_{51} + X_{24} X_{46} 
X_{62} \, - \nonumber  \\
&-& X_{23} X_{35} X_{56} X_{62}\, -\, X_{13} X_{34} X_{46} X_{61}
\, -\,  X_{12} X_{24} X_{45} X_{51} 
\eeqa
in self-explanatory notation.

A basis of fractional branes is given by the rank vectors
$(1,0,0,1,0,0)$, $(0,0,1,0,0,1)$  and $(1,0,1,0,1,0)$.

\subsection{The cascade for the first branch}

In this section we describe a cascading RG flow for the $dP_3$ theory. This duality cascade, which has not appeared in the 
literature, provides the dual of the throat in \cite{Franco:2004jz} corresponding to the appropriate choice of fractional branes.

The cone over $dP_3$ has a two-dimensional deformation branch, shown 
in Figure \ref{dp3}c, which involves two independent 
3-cycles and hence two independent RR fluxes. Hence a warped throat ending in this deformation 
must be dual to an RG flow in the quiver gauge theory with two independent fractional branes. From 
the geometry, and the argument in section \ref{topocons}, the 3-cycles involved in the deformation
correspond to the fractional branes with rank vectors $(1,0,0,1,0,0)$ and $(0,0,1,0,0,1)$.

Hence our starting point is the quiver in Figure \ref{quiverdp3} with ranks

\beq
\vec{N}=N (1,1,1,1,1,1)+P(1,0,0,1,0,0)+M(0,0,1,0,0,1)
\eeq

In addition, the $\IZ_2$ symmetry of the external legs in the toric diagram suggests that it is natural
to consider initial conditions such that the  RG flow is symmetric with respect to opposite nodes 
in the quiver. Hence, opposite nodes are taken with equal gauge couplings at a large UV scale. In 
order to study the RG flow to the infrared, we center on the
regime $N \gg P\gg M$, which eventually will lead to two hierarchically
different scales of RG flow. 

\begin{figure}
\begin{center}
\centering
\epsfysize=4.5cm
\leavevmode
\epsfbox{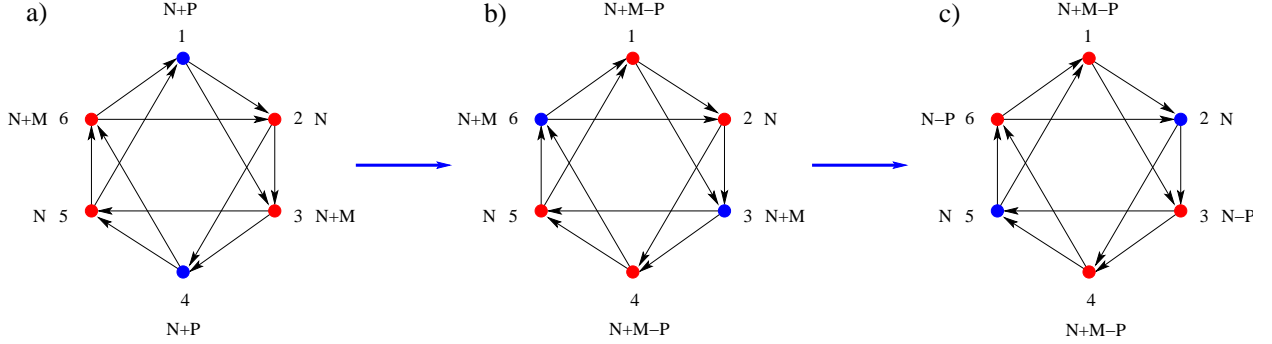}
\end{center}
\caption[]{\small Two dualizations in the first RG cascade in $dP_3$. Dualized nodes are shown in blue.}
\label{cascade1}
\end{figure}

The suggested duality cascade proceeds as follows. The nodes with largest beta function are 1 and 
4, so we dualize them simultaneously. The results are shown in Figure \ref{cascade1}a,b (the 
resulting quiver may be reordered to yield a standard maximally symmetric quiver, but we need 
not do so). Next the most strongly coupled nodes are 3, 6, so we dualize them simultaneously. This 
is shown in Figure \ref{cascade1}b,c.

The quiver in Figure \ref{cascade1}c can be reordered into a standard maximally symmetric quiver. 
This is of the form of the starting quiver, with similar 
fractional branes, but with the effective $N$ reduced by an amount $P$. We can then continue 
dualizing nodes 2, 5, then 1, 4, then 3, 6, etc, following the above 
pattern and generating a cascade 
which preserves the fractional branes but reduces the effective $N$.

In order to check that the suggested cascade of dualizations is consistent with an
RG flow, we compute the gauge theories at each step in the cascade, along with the beta
functions for the gauge couplings.

Given that $N \gg P \gg M$, we expect the cascade to be controlled by the $P$ fractional branes
of the first type in the UV. In that spirit, we study in detail the RG flow first neglecting the 
effect of $M$, which we set to zero for simplicity, under the assumption that the $M$ fractional 
branes of the second type will only produce a small perturbation to the cascade constructed this way.

Let us explore in more detail that the above proposed cascade of dualizations 1,4,2,5,3,6
indeed corresponds to an RG flow. This cascade iterates between Models I and II of $dP_3$ in 
\cite{Feng:2001bn}, and the corresponding ranks and beta functions at each step are

\beq
\begin{array}{l|cccccc|cccccc}
            &  \ \ N_1 \ \ &  \ \ N_2 \ \ &  \ \ N_3 \ \ &  \ \ N_4 \ \ &  \ \ N_5 \ \ &  \ \ N_6 \ \ &  \ \beta_1/P \ & \ \beta_2/P \ & \ \beta_3/P \ & \ \beta_4/P \ & \ \beta_5/P \ & \ \beta_6/P \ \\
\hline 
\hline 
{\bf 1} \ &    N+P &  N &    N &    N+P &    N &    N          & {\bf 3} & -3/2 & -3/2 & 3 & -3/2 & -3/2 \\
{\bf 2} \ &    N-P &  N &    N &    N+P &    N &    N           & -3 & 0 & 0 & {\bf 3} & 0 & 0 \\
{\bf 3} \ &    N-P &  N &    N &   N-P &  N &    N           & -3 & {\bf 3/2} & 3/2 & -3 & 3/2 & 3/2 \\
{\bf 4} \ &    N-P &  N-P &    N &   N-P &  N &    N           & -2 & -3/2 & 5/2 & -5/2 & {\bf 3/2} & 2 \\
{\bf 5} \ &    N-P &  N-P &    N &    N-P &  N-P &    N          & -3/2 & -3/2 & {\bf 3} & -3/2 & -3/2 & 3 \\
{\bf 6} \ &    N-P &  N-P &    N-2P &    N-P &  N-P &    N           & 0 & 0 & -3 & 0 & 0 & {\bf 3} \\
{\bf 7} \ & N-P &  N-P &    N-2P &    N-P &  N-P &    N-2P          & 3/2 & 3/2 & -3 & 3/2 & 3/2 & -3 
\end{array}
\eeq
After six dualizations (step ${\bf 7}$ in the previous table), the quiver comes back to itself, with 
ranks

\beq
\vec{N}=(N-P)(1,1,1,1,1,1)-P(0,0,1,0,0,1)
\eeq

Thus, the theory after six steps looks like the original one, with $N \rightarrow N-P$, plus a 
rotation and a replacement of $P \rightarrow -P$.

Notice that in the situation which is $\IZ_2$ symmetric with respect to 
opposite nodes, the above duality steps group by pairs of 
simultaneous dualizations, and the quivers involved are always maximally symmetric (model I in
\cite{Feng:2001bn}).

One may worry that in the presence of non-zero $M$ the structure of the above cascade is destabilized. 
However, numerical results on the structure of cascades for a variety of choices of UV gauge couplings 
shows that the existence of cascades is a quite robust feature of the above choice of fractional 
branes (although the particular pattern of dualities involved in a cycle may be different from the 
above one). 

Hence, the above cascade can be generalized to the situation with non-zero $M$, with the same result, 
namely there are cycles of Seiberg dualities, which leave the quiver and fractional branes 
invariant, but decrease the number of D3-branes in multiples of $P$.

\begin{figure}
\begin{center}
\centering
\epsfysize=5cm
\leavevmode
\epsfbox{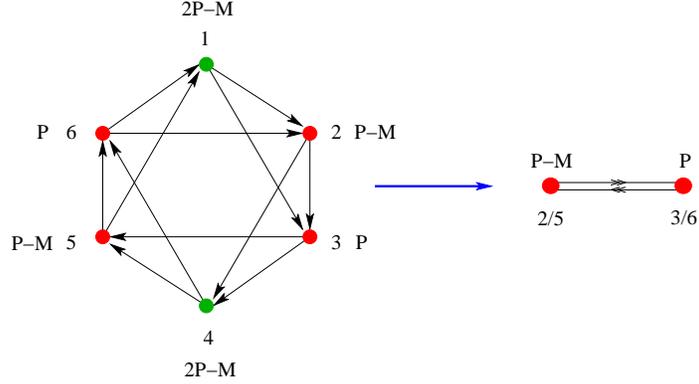}
\end{center}
\caption[]{\small Condensation of the gauge theory of $dP_3$ to the
gauge theory of the conifold. The nodes undergoing a deformation are indicated in green.}
\label{condensation}
\end{figure}

The cascade proceeds until the effective $N$ is not large compared with $P$. 
For simplicity, consider that the starting $N$ is $N=(k+2)P-M$. Then 
after a suitable number of cascade steps, the ranks in the maximally 
symmetric quiver are $(2P-M, P-M, P, 2P-M, P-M, P)$, for nodes 
$(1,2,3,4,5,6)$, as shown in \fref{condensation}a. At this stage the $SU(2P-M)$ factors 
have $2P-M$ flavors and develop a quantum deformation of their moduli space. This should 
correspond to turning on one of the complex deformations of the geometry.  
From the structure of the left over web in the toric representation after a 
one-parameter deformation, see Figure \ref{dp3}b, we expect that the left
over geometry should be a conifold. This is shown 
in Figure \ref{condensation}.

Before describing this quantum deformation in detail, let us simply mention that it results in the 
disappearance of nodes 1 and 4, the recombination of nodes 2 and 3, and
5 and 6 respectively, due to meson vevs, and a rearrangement of the arrows. The final result is 
indeed a conifold quiver gauge theory, with ranks $P-M$ and $P$. The theory subsequently evolves 
towards the infrared via a Klebanov-Strassler flow, by duality cascades where the effective number 
of D3-branes decreases in steps of $M$. At the end of this cascade, there is another condensation, 
which corresponds to turning on the second complex deformation of the cone over $dP_3$ to yield a 
smooth space.

\subsection{The quantum deformation to the conifold}

\label{section_dP3_to_conifold}

\begin{figure}
\begin{center}
\centering
\epsfysize=4.5cm
\leavevmode
\epsfbox{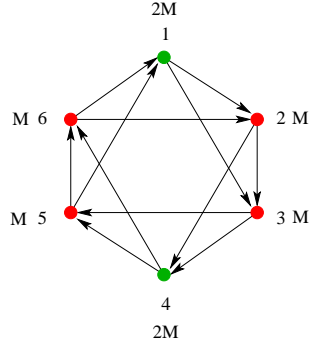}
\end{center}
\caption[]{\small Gauge theory encoding the dynamics of D3-brane probes of the infrared of the 
cascade. The nodes undergoing a deformation are indicated in green.}
\label{laststep}
\end{figure}

Let us now describe the fate of the $dP_3$ quiver theory at the end of 
the first
duality cascade. To simplify the discussion, we take the situation where nodes 2356 have equal 
rank, i.e. $M=0$, but the generalization to non-zero $M$ is possible. We 
would like to consider the gauge 
theory associated to a set of $P$ D3-branes probing the infrared of the duality cascade.
The corresponding quiver is shown in Figure \ref{laststep}. 

Following our general discussion in section \ref{generaldeform}, the $SU(2P)$ nodes condense, so
we introduce the corresponding mesons
\beqa
{\cal M}=\left[\begin{array}{cc} M_{63} & M_{62} \\ M_{53} & M_{52} \end{array}\right] =
\left[\begin{array}{cc} X_{61} X_{13} & X_{61} X_{12} \\ X_{51} X_{13} & X_{51} X_{12}
\end{array}\right]
\quad ; \quad
{\cal N}=
\left[\begin{array}{cc} N_{36} & N_{35} \cr N_{26} & N_{25} \end{array}\right]=
\left[\begin{array}{cc} X_{34} X_{46} & X_{34} X_{45} \\ X_{24} X_{46} & X_{24} X_{45}
\end{array}\right] 
\nonumber
\eeqa
We also introduce the baryons ${\mathcal B}$, ${\tilde {\mathcal B}}$, ${\mathcal A}$, ${\tilde 
{\mathcal A}}$. The
quantum constraints read
\beqa
\det {\mathcal M} - {\mathcal B}{\tilde {\mathcal B}}=\Lambda^{4P} \quad ; \quad
\det {\mathcal N} - {\mathcal A}{\tilde {\mathcal A}}=\Lambda^{4P}
\eeqa
where we use the same dynamical scale for both gauge groups, corresponding to the $\IZ_2$ symmetry
of opposite nodes in the quiver preserved during the flow.

The superpotential reads
\beqa
W & = & M_{62} X_{23} N_{35} X_{56}\, +\, M_{53} X_{35} + N_{26} X_{62}\, - \nonumber  \\
&-& X_{23} X_{35} X_{56} X_{62}\, -\, M_{63} N_{36}\, -\, M_{52} N_{25}+ \nonumber \\
&+& X_1\, (\det {\mathcal M} - {\mathcal B}{\tilde {\mathcal B}}-\Lambda^{4P})\, 
+\, X_2\, (\det {\mathcal N} - {\mathcal A}{\tilde {\mathcal A}}-\Lambda^{4P})
\eeqa

Going along the mesonic branch, we uncover the dynamics of the probes in the geometry at the 
infrared of the cascade. The mesonic branch corresponds to
\beqa
X_1=X_2=\Lambda^{4-4P} \quad ; \quad {\mathcal A}={\tilde {\mathcal 
A}}=0\quad ; \quad {\mathcal B}={\tilde 
{\mathcal B}}=0
\eeqa
and the constraints on the mesons. For the most symmetric choice of meson vevs ${\cal 
M}\propto\id$, ${\cal N}\propto \id$, the gauge groups associated to the nodes 3 and 6, and 2 and 
5, are broken to their respective diagonal combinations.

In order to simplify the discussion, we restrict to the Abelian case, where the superpotential 
reads
\beqa
W & = & M_{62} X_{23} N_{35} X_{56} -  X_{23} X_{35} X_{56} X_{62} - M_{63} N_{36}
- M_{52} N_{25}+ \nonumber \\
& + & M_{53} X_{35} + N_{26} X_{62}+ M_{63}M_{52}-M_{53}M_{62} +N_{36}N_{25}-N_{26}N_{35} 
\label{W_dP3_mesonic}
\eeqa
Using the equations of motion for $M_{53}$ and $N_{26}$, we have $X_{35}=M_{62}$, 
$X_{62}=N_{35}$. Thus
\beqa
W & = & X_{23} N_{35} X_{56} M_{62} - X_{23} M_{62} X_{56} N_{35} - \nonumber \\
& -& M_{63} N_{36} - M_{52} N_{25}+M_{63}M_{52}+N_{36}N_{25}
\label{almosthere}
\eeqa
Using the equations of motion for e.g. $M_{63}, M_{52}$, the quadratic terms disappear, and we are 
left with
\beqa
W & = & X_{23} N_{35} X_{56} M_{62} - X_{23} M_{62} X_{56} N_{35} 
\eeqa
Going back to the non-Abelian case, the gauge group is $SU(M)_{25}\times 
SU(M)_{36}$, with charged fields given by those appearing in the  
superpotential. These can be relabeled as $A_1=X_{23}$ and $A_2=X_{56}$, 
in the $(\fund,\antifund)$, and $B_1=M_{35}$, $B_2=M_{62}$, in the 
$(\antifund,\fund)$. This is the gauge theory of D3-branes at a 
conifold singularity, showing that the left over geometry after the 
complex deformation is a conifold. It is important to note that there 
are some additional massless fields, which describe the dynamics of the 
D3-brane probe in the deformed geometry.
Specifically, the quadratic terms in (\ref{almosthere}) leave two linear combinations of $M_{63}$, 
$M_{52}$, $N_{36}$, $N_{25}$ massless. In addition, the fields $M_{53}$ and $N_{26}$, which 
disappeared from the superpotential, also remain massless. Overall, we have light fields
subject to the constraints (from $\partial W/\partial X_i=0$)
\beqa
M_{63}M_{52}-M_{53}M_{62}=\Lambda^{4P}\quad ; \quad N_{36}N_{25}-N_{26}N_{35} =\Lambda^{4P}
\eeqa
Hence the complete dynamics of the theory corresponds to one D3-probe in a geometry which is the 
deformation of a complex cone over $dP_3$ to a singular conifold.

Notice also that if we consider two kinds of fractional branes, namely non-zero $M$ in the 
original cascade, the quantum deformation proceeds as above, since it involves recombinations of 
opposite nodes which have equal ranks even for non-zero $M$. The resulting condensation leads to a 
conifold, with the two nodes of the conifold theory having different ranks, what triggers a further 
Klebanov-Strassler duality cascade and infrared deformation. 

\subsection{The other branch}

The cone over $dP_3$ has a second deformation branch, which is one-dimensional, see figure 
\ref{dp3}c. In this section we discuss the duality cascade dual to the corresponding supergravity
throat, and describe the infrared deformation in the gauge theory.

\begin{figure}
\begin{center}
\centering
\epsfysize=4.5cm
\leavevmode
\epsfbox{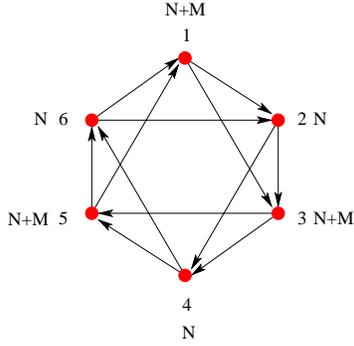}
\end{center}
\caption[]{\small Starting point of the cascade ending in the
one-parameter deformation of the cone over $dP_3$.}
\label{secondbranch}
\end{figure}

Using the relation in section \ref{topocons}, the one-parameter deformation branch corresponds to the 
choice of fractional branes in Figure \ref{secondbranch}. Also, due to the $\IZ_3$ symmetry of the 
geometry, it is natural to propose that nodes with even/odd label have equal UV 
couplings, respectively.

\begin{figure}
\begin{center}
\centering
\epsfysize=4.5cm
\leavevmode
\epsfbox{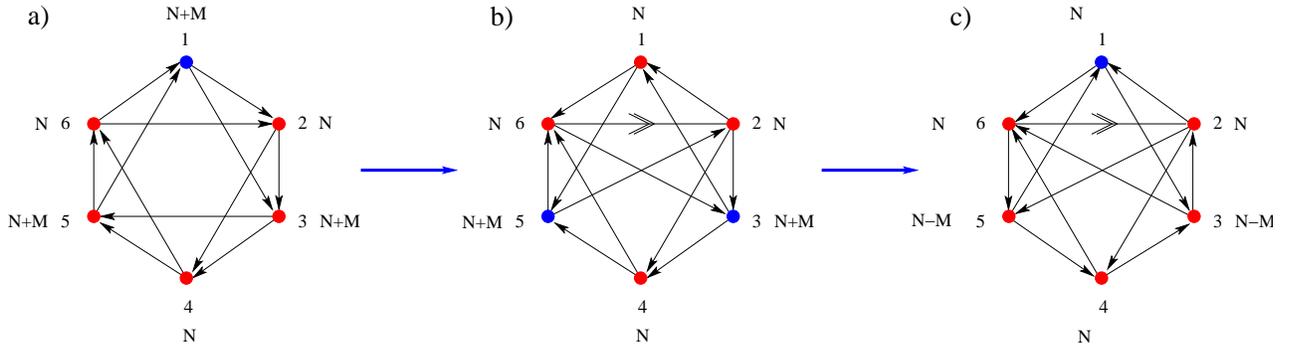}
\end{center}
\caption[]{\small Some steps in the duality cascade. Dualized nodes are shown in blue.}
\label{cascade2}
\end{figure}

The proposed cascade in this case goes as follows. As one flows to the infrared, 
the $SU(N+M)$ gauge factors become strongly coupled and should be dualized.
Their simultaneous dualization is difficult, since there are bi-fundamentals
joining the corresponding nodes, so we proceed sequentially, with a particular choice of ordering
which is not important for the final result. We choose to dualize node 1 first. The result is shown in 
Figure \ref{cascade2}ab. In the resulting theory, there are no bi-fundamentals joining nodes 3 and 5, so 
we can now dualize them simultaneously, as shown in Figure \ref{cascade2}bc.

\begin{figure}
\begin{center}
\centering
\epsfysize=4.5cm
\leavevmode
\epsfbox{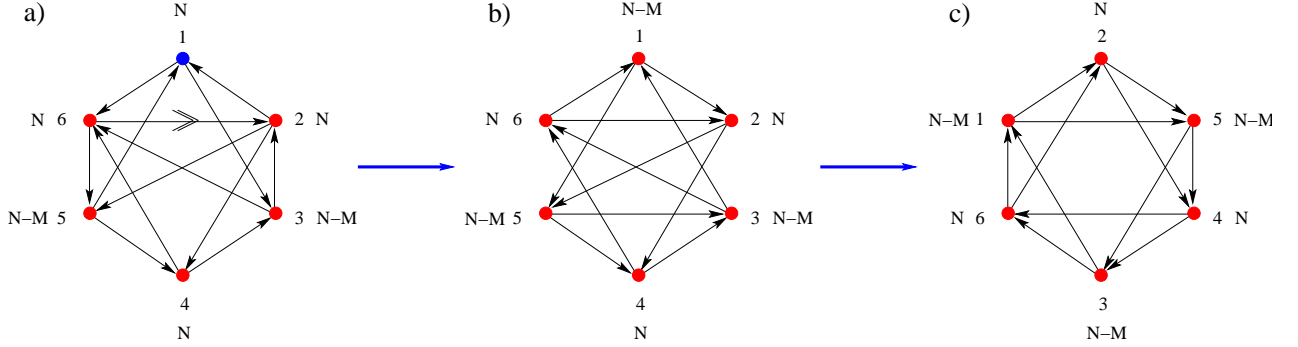}
\end{center}
\caption[]{\small Last duality and reordering to complete the duality step. Note that the nodes of the last quiver have been reordered
in order to make its $\IZ_3$ symmetry manifest. The dualized node is shown in blue.}
\label{final2}
\end{figure}

Next, node 1 is most strongly coupled, so we dualize it again. The result is shown in Figure \ref{final2}ab. 
Then, we dualize nodes 2 and 6. The final quiver is the maximally symmetric one, as can be shown by reordering the nodes as in
\ref{final2}bc. This final theory is of the same kind as the original one, but reducing the effective 
$N$ in $M$ units (and up to a rotation). Notice also that the final theory has the same nice $\IZ_3$ 
symmetry between the nodes as the original onem with the nodes $(2,4,6)$ playing the role of $(1,3,5)$. One can then proceed to 
perform the same sequence of dualizations, this time on nodes $(2,4,6)$, completing a full cycle of the cascade.

The above heuristic derivation is confirmed by the detailed 
computation, and provides 
the field theory interpretation of the supergravity solution 
in \cite{Franco:2004jz}, for the corresponding choice of asymptotic fluxes. The cascade proceeds until 
the effective number of D3-branes is comparable to that of fractional branes. At this stage, we may 
use the field theory to derive the strong infrared dynamics which 
removes the singularity by 
replacing it by a smooth deformed geometry.

For that purpose, we consider the dynamics of the theory at the end of the 
cascade, in the presence of additional D3-brane probes. Namely, we 
consider the quiver with ranks $(2M,M,2M,M,2M,M)$. In this situation, we 
expect that the three 
nodes 1, 3, 5 lead to a quantum deformed moduli space. In order to study the left-over theory, we 
consider performing this condensations sequentially (the order not being relevant for the 
final result).

Consider the strong dynamics associated to the node 1. We introduce the mesons 
\beqa
{\cal M}=\left[\begin{array}{cc} M_{62} & M_{63} \\ M_{52} & M_{53} \end{array}\right] =
\left[\begin{array}{cc} X_{61} X_{12} \, & \, X_{61} X_{13} \\ X_{51} X_{12}\,  &\, X_{51} X_{13}
\end{array}\right]
\eeqa
Similar to our above analysis, we implement the quantum constraint in the superpotential. We 
center on the mesonic branch, along which the gauge factors 6 and 2, and 5 and 3, are broken to 
their respective diagonal subgroups, denoted 26 and 35 henceforth. 
Restricting to the Abelian case, the superpotential is described by
\beqa
W & = & M_{62} X_{23} X_{34} X_{45} X_{56} \, + \, M_{53} X_{35}
\, + \, X_{24} X_{46} X_{62} \, -  \, X_{23} X_{35} X_{56} X_{62} \nonumber \\
&-&  M_{63} X_{34} X_{46} \, - M_{52} X_{24} X_{45} \,- \, 
M_{62}M_{53} \, + \, M_{52} M_{63}
\eeqa
The combined node 35 has $N_f=N_c$ plus additional massive adjoints and flavors, which we integrate 
out using the equations of motion for $M_{53}$, $X_{35}$, $M_{52}$, $M_{63}$. The resulting 
superpotential is
\beqa
W & = & M_{62} X_{23} X_{34} X_{45} X_{56} \, + \, X_{24} X_{46} X_{62} \, - \, 
-\, X_{23} M_{62} X_{56} X_{62} \, - \, X_{34} X_{46} X_{24} X_{45} 
\eeqa
so the only fields charged under the node 35 are the massless ones. Since it has
$N_f=N_c$ we introduce the mesons
\beqa
{\cal N}=\left[\begin{array}{cc} N_{26} & N_{24} \\ N_{46} & N_{44} \end{array}\right] =
\left[\begin{array}{cc} X_{23} X_{56}\, &\, X_{23} X_{34} \\ X_{45} X_{56}\, &\, X_{45} X_{34}
\end{array}\right]
\eeqa
which, along with the corresponding baryons, satisfy a quantum deformed constraint.
Along the mesonic branch, the group associated to the nodes 26 and 4 is broken to a single 
diagonal combination. The superpotential is given by
\beqa
W & = & M_{62} N_{24} N_{46} \, + \, X_{24} X_{46} X_{62} \, - \, 
N_{26} M_{62} X_{62} \, -\, N_{44} X_{46} X_{24} \, + \, N_{26}N_{44} \, 
-\, N_{46}N_{24}\quad
\eeqa

Using the equations of motion for $N_{26}$, $N_{44}$, $N_{46}$, $N_{24}$, the superpotential reads
\beqa
W & = & X_{24} X_{46} X_{62} \, - \, X_{24} X_{62} X_{46}
\eeqa
Since these fields transform in the adjoint representation of the leftover $SU(M)$ gauge group, this is the ${\cal N}=4$ SYM theory, 
and the result implies that the 
geometry after the deformation is smooth. As usual, there are some additional neutral massless fields, with quantum 
modified constraints, which describe the dynamics of the probe in the deformation of $dP_3$ to 
a smooth geometry. We see that the complete smoothing by a single scale is in full
agreement with the geometric picture.

\section{Further examples}
\label{further}

In this section we apply our by now familiar techniques to study other 
examples of quiver gauge theories with two scales of strong infrared 
dynamics.

\subsection{From $PdP_4^{(I)}$ to the Suspended Pinch Point}

We now investigate a two-scale cascade which follows the sequence

\beq
PdP_4^{(I)} \rightarrow SPP \rightarrow {\rm smooth} \nonumber
\eeq
where $PdP$ stands for 'pseudo del Pezzo' and $PdP_4^{(I)}$ indicates the complex cone 
over a non-generic toric blow-up of $dP_3$ denoted Model I of $PdP_4$ in \cite{Feng:2002fv}.

This is another simple example of the agreement between the complex deformation of the geometry, and the quantum
deformation of D3-branes probing the infrared theory of fractional branes. Since the discussion of the RG flow
and existence of cascades in these geometries is involved and somewhat aside our main interest, we skip their discussion and
center on the gauge theory description of the deformation.

\begin{figure}[ht]
  \epsfxsize = 11cm
  \centerline{\epsfbox{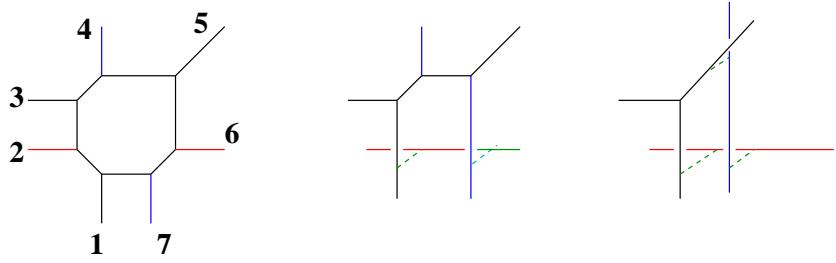}}
  \caption{Web diagram for the $PdP_4^{(I)}$ model, its deformation to 
the SPP, and a further deformation to a smooth space.}
  \label{web_PdP4I}
\end{figure}

We consider the theory on a stack of D3-branes probing a complex cone over the toric variety obtained
by performing a non-generic blow-up of $dP_3$. Figure \fref{web_PdP4I}a shows the $(p,q)$ web diagram for this geometry. We also 
indicate a complex deformation to the suspended pinch point (SPP) singularity.

\begin{figure}[ht]
  \epsfxsize = 6.5cm
  \centerline{\epsfbox{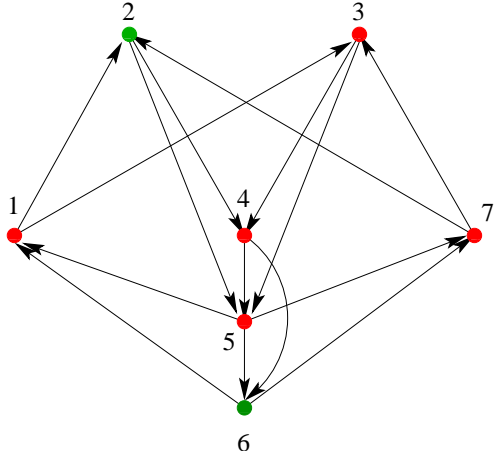}}
  \caption{Quiver diagram for $PdP_4^{(I)}$. We show in green the nodes that undergo the deformation.}
  \label{quiver_PdP4}
\end{figure}

The quiver diagram for this model is shown in \fref{quiver_PdP4}, which has a 5-block structure that is 
evident in the web diagram, with nodes 7, 1 and 2, 3 forming pairs.

The corresponding superpotential was derived in \cite{Feng:2002fv} and 
reads

\beq
\begin{array}{rl}
W & =X_{24}X_{46}X_{61}X_{12}+X_{73}X_{35}X_{57}-X_{73}X_{34}X_{46}X_{67}
-X_{45}X_{57}X_{72}X_{24} \\
  & -X_{35}X_{56}X_{61}X_{13}+X_{51}X_{13}X_{34}X_{45}-X_{25}X_{51}X_{12}+X_{25}X_{56}X_{67}X_{72}
\end{array}
\eeq

Following our arguments in section \ref{topocons}, the deformation we want to consider corresponds, in the gauge theory, to the 
choice of fractional branes

\beq
\vec{N}=M(1,1,1,1,1,1,1)+M(0,1,0,0,0,1,0)
\eeq

Following our general prescription, we construct the meson fields for nodes 2 and 6

\begin{eqnarray}
\mathcal{M} = \left[\begin{array}{cc} M_{14} & M_{74} \\  M_{15} & M_{75} \end{array} \right]
            = \left[\begin{array}{ccc} X_{12}X_{24} & \ & X_{72}X_{24} \\ X_{12}X_{25} & \ & 
X_{72}X_{25} \end{array} \right] \nonumber \\ \\
\mathcal{N} = \left[\begin{array}{cc} N_{41} & N_{47} \\  N_{51} & N_{57} \end{array} \right]
            = \left[\begin{array}{ccc} X_{46}X_{61} & \ & X_{46}X_{67} \\ X_{56}X_{61} & \ & 
X_{56}X_{67} \end{array} \right] \nonumber
\end{eqnarray}

We now introduce Lagrange multiplier chiral fields to impose the quantum modified constraints on 
mesons and baryons. Along the mesonic branch we have
\beqa
& X_1=\Lambda^{4-4M} \quad ; \quad \mathcal{B}=\tilde{\mathcal{B}}=0 \quad 
; \quad 
X_2=\Lambda^{4-4M} \quad ; \quad \mathcal{C}=\tilde{\mathcal{C}}=0 & 
\nonumber \\
& \det\mathcal{M}=\Lambda^{4M}\quad ; \quad  \det\mathcal{N}=\Lambda^{4M}
\eeqa
Along the mesonic branch, nodes 1, 4 and 5, 7, recombine to their respective diagonal 
combinations. Restricting to the Abelian case, the superpotential is

\beq
\begin{array}{rl}
W & 
=M_{14}N_{41}+X_{73}X_{35}X_{57}-
X_{73}X_{34}N_{47}-X_{45}X_{57}M_{74} \\
  & 
-X_{35}N_{51}X_{13}+X_{51}X_{13}X_{34}X_{45}-
M_{15}X_{51}+M_{75}N_{57} \\
  & - M_{14}M_{75}+ M_{15}M_{74} -N_{41}N_{57} + N_{51}N_{47}
\end{array}
\eeq

Using the equations of motion for $M_{14}$, $M_{15}$, $N_{57}$, $N_{47}$, 
$N_{51}$, etc, we have
\beqa
N_{41}=M_{75} \quad ; \quad X_{51}=M_{74} \quad ; \quad M_{75}=N_{41} \quad N_{51}=X_{73}X_{34}
\quad ; \quad N_{47}=X_{35}X_{13}
\eeqa
The gauge group after symmetry breaking is $SU(N)_{57}\times SU(N)_{14}\times SU(N)_{3}$, and we have the superpotential
\beq
\begin{array}{rl}
W & 
=X_{73}X_{35}X_{57}-M_{74}X_{45}X_{57}-X_{73}X_{34}N_{47} 
-X_{35}X_{73}X_{34}X_{13}+M_{74}X_{13}X_{34}X_{45}
\end{array}
\eeq
Relabeling the gauge group as $SU(N)_1\times SU(N)_2\times SU(N)_3$, and the fields as
\beqa
& M_{74}\to Y_{12} \quad , \quad X_{45}\to Y_{21} \quad , \quad X_{13}\to Y_{23} \quad , \quad X_{34} \to Y_{32}
\nonumber \\
& X_{35}\to Y_{31}\quad , \quad X_{71}\to Y_{13} \quad , \quad X_{57}\to Y_{11} &
\eeqa
we readily see the field content and superpotential of the SPP geometry. In addition to these 
fields, there are some massless modes, left over from the initial mesons. One can check that out 
of the eight original fields, five combinations remain massless, and they are subject to the 
quantum constraints, hence three degrees of freedom remain. They provide the moduli space of a 
D3-brane probe in the geometry given by the $dP_4$ deformed to a SPP.

The remaining theory may have fractional branes, triggering an RG flow related to the sequence of dualities discussed for SPP in section \ref{thespp}, which terminates in smooth $\IC^3$.

\subsection{From $PdP_3b$ to $\IC²/\IZ_2$}

In this section we would like to discuss a further example of 
condensation, realized geometrically as the deformation of 
a non-generic blow-up of $dP_2$, the pseudo del Pezzo denoted $PdP_3b$ in 
\cite{Feng:2004uq}, to a $\IC^2/\IZ_2\times \IC$ orbifold singularity. From 
the geometric viewpoint, it illustrates the fact that different phases of the 
quiver gauge theory may suffer different condensation 
processes. From the field-theoretical viewpoint, it provides an example 
with a different behavior for the left over 
theory. Namely, instead of the $\mathcal{N}=4$ theory or a conifold-like 
singularity, the left-over geometry corresponds to an 
orbifold singularity. In the presence of fractional branes on 
$\IC^2/\IZ_2$, the theory is not conformal, but instead of 
running down a cascade it encounters a singularity. The smoothing of this 
singularity in the dual supergravity description is of enhan\c{c}on type 
\cite{Johnson:1999qt}.

\medskip

\begin{figure}[ht]
  \epsfxsize = 4cm
  \centerline{\epsfbox{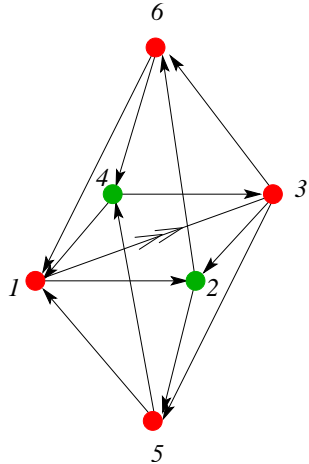}}
  \caption{Quiver diagram for $PdP_3b$.}
  \label{quiver_dP3_2}
\end{figure}

\begin{figure}[ht]
  \epsfxsize = 8cm
  \centerline{\epsfbox{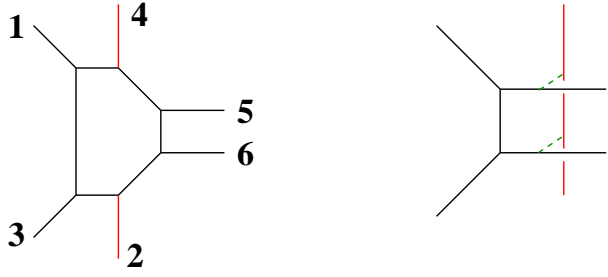}}
  \caption{Web diagram for the cone over the non-generic blow-up of $dP_2$, and its deformation.
The external legs have been labeled indicating their correspondence to the nodes in the quiver
in \fref{quiver_dP3_2}.}
  \label{pdp3}
\end{figure}

Let us consider a set of branes at a complex cone over the non-generic blow-up of $dP_2$
leading to the quiver gauge theory in the phase denoted Model II of $PdP_3b$, worked out in \cite{Feng:2004uq}, and 
whose quiver diagram is shown in \fref{quiver_dP3_2}. The corresponding 
toric web diagram is shown in Figure \ref{pdp3}
\footnote{Here we adhere to the terminology introduced in \cite{Feng:2004uq}.
Thus, we see that the toric diagram for $PdP_3b$ is different from the one for $dP_3$
and is given by the reciprocal of the $(p,q)$ web in \fref{pdp3}.}.

The tree level superpotential 
is given by
\beq
\begin{array}{rl}
W_0=&X_{12}X_{25}X_{54}X_{41}+X_{26}X_{64}X_{43}X_{32}-X_{25}X_{51}Y_{13}X_{32}-X_{64}X_{41}X_{13}X_{36} \\
    &+Y_{13}X_{36}X_{61}+X_{13}X_{35}X_{51}-X_{61}X_{12}X_{26}-X_{43}X_{35}X_{54}
\end{array}
\eeq

The geometric deformation of this space is shown in Figure \ref{pdp3}b. Using our arguments in section \ref{topocons}, this 
corresponds to strong coupling dynamics associated to nodes 2 and 4 in the quiver diagram. In order to show this using D3-brane 
probes of this infrared dynamics, we consider the quiver gauge theory with rank vector 
\beq
\vec{N}=M (0,1,0,1,0,0)+M (1,1,1,1,1,1)=M(1,2,1,2,1,1)
\eeq

In this situation, the nodes 2 and 4 have $N_f=N_c$, and have a quantum deformed moduli space. Hence the above gauge 
theory (along the mesonic branch) describes the dynamics of D3-brane probes in the left over geometry after the complex 
structure deformation of the original geometry $PdP_3$. In the following, we follow the by now familiar arguments to 
determine the latter.

We introduce the meson fields 
\begin{eqnarray}
\mathcal{M} = \left[\begin{array}{cc} M_{15} & M_{35} \\  M_{16} & M_{36} \end{array} \right]
            = \left[\begin{array}{ccc} X_{12}X_{25} & \ & X_{32}X_{25} \\ X_{12}X_{26} & \ & X_{32}X_{26} \end{array} \right] \nonumber \\ \\ 
\mathcal{N} = \left[\begin{array}{cc} N_{51} & N_{53} \\  N_{61} & N_{63} \end{array} \right]  
            = \left[\begin{array}{ccc} X_{54}X_{41} & \ & X_{54}X_{43} \\ X_{64}X_{41} & \ & X_{64}X_{43} \end{array} \right] \nonumber
\end{eqnarray}

In terms of mesons and baryons, the superpotential becomes

\beq
\begin{array}{rl}
W=&M_{15} N_{51}\, + \, M_{36} N_{63}\, -\, M_{35}X_{51}Y_{13}\, -\, N_{61}X_{13}X_{36} \\
  &+\, Y_{13}X_{36}X_{61}\, +\, X_{13}X_{35}X_{51}\, -\, M_{16}X_{61} \, -\, N_{53}X_{35} \\
  &-X_1\, (\det\mathcal{M}-\mathcal{B}\tilde{\mathcal{B}}-\Lambda^{4M})-X_2 \,
(\det\mathcal{N}-\mathcal{C}\tilde{\mathcal{C}}-\Lambda^{4M})
\end{array}
\label{W_example_1}
\eeq
The mesonic branch is given by
\beqa
X_1=X_2=\Lambda^{4-4M} \quad ; \quad 
\mathcal{B}=\tilde{\mathcal{B}}=0\quad 
;\quad \mathcal{C}=\tilde{\mathcal{C}}=0
\eeqa
with the mesons subject to the quantum constraints. Also, along the mesonic branch, the symmetry is broken by recombining 
the gauge factors 1 and 5, and 3 and 6, into their respective diagonal combinations. 

Restricting now to the Abelian case, the superpotential is

\beq
\begin{array}{rl}
W=&M_{15} N_{51} \, +\, M_{36} N_{63}\, -\, M_{35}X_{51}Y_{13}\, -\, N_{61}X_{13}X_{36} \\
  &+\, Y_{13}X_{36}X_{61} \, +\, X_{13}X_{35}X_{51}\, -\, M_{16}X_{61}\, -\, N_{53}X_{35} \\
  &-\, M_{15} M_{36} \, +\, M_{16} M_{35} \, -\, N_{51} N_{63} \, +\, 
N_{61} N_{53}
\end{array}
\eeq

\begin{figure}[ht]
  \epsfxsize = 4cm
  \centerline{\epsfbox{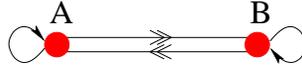}}
  \caption{Quiver diagram after deformation of $PdP_3b$. It corresponds to a $\IC^2/\IZ_2\times \IC$ geometry.}
  \label{quiver_deformed}
\end{figure}

Using the equations of motion, we obtain e.g. $N_{61}=X_{35}$,
$M_{16}=X_{61}$, $N_{51}=M_{36}$,
$M_{15}=N_{63}$. The superpotential is
\beqa
W & = & M_{35} X_{51} Y_{13}\, -\, N_{61} X_{13} X_{36}\,
+\, Y_{13} X_{36} M_{16}\, +\, X_{13} N_{61}X_{51}
\eeqa
Relabeling the unbroken group as $SU(N)_A\times SU(N)_B$, and the fields as $Y_{13}\to X_{AB}$, $M_{35}\to X_{BA}$, 
$X_{51}\to \Phi_{AA}$, $X_{13}\to Y_{AB}$, $N_{61}\to Y_{BA}$, $X_{36}\to \Phi_{BB}$,  
the final quiver is presented in \fref{quiver_deformed}.

The field content and superpotential correspond to the gauge theory for a
$\IC^2/\IZ_2\times \IC$ geometry. This agrees with the expected left over geometry after the complex deformation. In addition, the 
theory contains massless meson degrees of freedom, subject to the quantum constraint. They describe the dynamics of the D3-brane 
probe in the geometry given by the complex deformation of $PdP_3$ to $\IC^2/\IZ_2\times \IC$.

As usual, it is possible to study the situation where the final gauge theory 
contains fractional branes. This theory is $\mathcal{N}=2$ supersymmetric, hence its RG evolution 
could be determined from its exact solution. As usual in non-conformal $\mathcal{N}=2$ theories, 
instead of a duality cascade we expect strong coupling singularities. In the dual supergravity side, 
they are described as enhan\c{c}on configurations \cite{Johnson:1999qt}.

\section{Conclusions}
\label{conclusions}

In this paper we have centered on the gauge field theory dynamics 
associated to the smoothing of singularities in warped throat solutions 
dual to RG flows for branes at singularities in the presence of 
fractional branes. We have established that in a large set of examples the 
smoothing corresponds to a complex deformation of the cone
geometries. We have described this phenomenon in the dual gauge field 
theory, by using D3-brane probes of the infrared dynamics. The geometric 
deformation arises as a quantum deformation of the moduli space of the 
D3-brane probes. The field theory description is in full agreement with 
the geometric description of the complex deformation using toric methods.

In addition, we have constructed new explicit examples of cascading RG 
flows for some of these theories. These duality cascades, along with the 
infrared deformations, are generalizations of the Klebanov-Strassler RG 
flow, but show a richer structure in several respects. For instance, very 
interestingly, several examples correspond to duality cascades with several
scales of partial confinement and deformation, after each of which the 
remaining quiver theory continues cascading down the infrared in a 
different pattern. Their supergravity duals should correspond to warped 
throats whose warp factor and flux structure jumps at particular values 
of the radial coordinate. In other words, to warped throats based on a 
deformed geometry with several 3-cycles, which are of hierarchically 
different size. It would be interesting to develop a better understanding of
these throats directly from the supergravity side. Also, we expect 
several interesting applications of these richer throat structures to 
compactification and model building \cite{csu}.

Our work opens a set of new questions. For instance, certain geometries do admit fractional branes, and even have known 
KT-like warped throat solutions, but do not admit complex deformations to smooth out their singularities. It would be 
interesting to understand the infrared behavior of this class of models. In particular, the real 
cones over the recently studied $Y^{p,q}$ manifolds, of which the five-dimensional horizon of the 
complex cone over $dP_1$ is an example, fall in this class. We hope interesting progress in this direction.

Finally, there is an interesting phenomenon taking place in the quiver 
gauge theories we have studied, which is however not involved in the nice 
RG flows we have centered on. Namely, some of these theories, for other 
choices of fractional branes (or of UV gauge couplings) exhibit duality 
walls \cite{Hanany:2003xh,Franco:2003ja,Franco:2003ea,Franco:2004jz}. It 
is conceivable that a gauge theory with in principle a duality wall in 
its UV can actually be UV completed by regarding it as a remnant after 
confinement of a larger gauge theory at higher energies, with a better 
behaved UV regime. Thus our work may shed some light also into these more 
exotic RG flows. We leave this and other questions for future 
research.

\section*{Acknowledgements}

We thank J.F.G. Cascales, Y. H. He, Y. Oz and F. Saad, for useful discussions. 
A. H. wishes to acknowledge the kind hospitality of the physics 
department at the Hebrew University, and The Fields Institute for Research 
in Mathematical Sciences where parts of this work were completed. S. F. 
would like to thank Y. H. He, C. Herzog and J. Walcher for enjoyable 
collaboration in previous related projects. A. M. U. thanks M. Gonz\'alez 
for encouragement and support.
The research of S. F. and A. H. was supported in part by the CTP and LNS of
MIT and the U.S. Department of Energy under cooperative research 
agreement $\#$ DE--FC02--
94ER40818, and by BSF American--Israeli Bi--National Science Foundation. A. H. is also
indebted to a DOE OJI Award. The research of A. M. U. was supported by the 
CICYT, Spain, under project FPA2003-02877.

\appendix
\section{A more careful look at the mesonic branch}
\label{mesonic}

In this appendix we present an alternative approach to the field theory analysis of the IR
complex deformation of the geometry, which complements our methods in Section \ref{generaldeform}.
The strategy will be to consider the dynamics of the fluctuations of the meson fields around
the expectation values required by the quantum constraints. As we will see, this method has the advantage
of clarifying how the relative signs of the Lagrange multipliers are determined and shows how
the low energy limit with respect to the strong coupling scales is taken explicitly.

In order to illustrate these ideas, we will focus in the example of the deformation from $dP_3$ down
to the conifold. We will reproduce the computations performed in Section \ref{section_dP3_to_conifold}
from a different viewpoint. 

As discussed, the quantum modified constraints on the meson and baryon 
fields \eref{qmodif} are imposed via
Lagrange multipliers $X_i$. The quiver for the phase of $dP_3$ we are considering is shown in
\fref{laststep}. The ranks are

\beq
\vec{N}=M(1,1,1,1,1,1)+M(1,0,0,1,0,0)
\eeq 
leading to a quantum modified moduli space for nodes 1 and 4. The meson fields
for these nodes are

\beqa
{\cal M}=\left[\begin{array}{cc} M_{63} & M_{62} \\ M_{53} & M_{52} \end{array}\right] =
\left[\begin{array}{cc} X_{61} X_{13} & X_{61} X_{12} \\ X_{51} X_{13} & X_{51} X_{12}
\end{array}\right]
\quad ; \quad
{\cal N}=
\left[\begin{array}{cc} N_{36} & N_{35} \\ N_{26} & N_{25} \end{array}\right]=
\left[\begin{array}{cc} X_{34} X_{46} & X_{34} X_{45} \\ X_{24} X_{46} & X_{24} X_{45}
\end{array}\right] \nonumber
\eeqa

In terms of them and the baryonic operators, the quantum corrected superpotential is

\beqa
W & = & M_{62} X_{23} N_{35} X_{56} -  X_{23} X_{35} X_{56} X_{62} - M_{63} N_{36} - M_{52} N_{25}+M_{53} X_{35} + N_{26} X_{62} 
\nonumber \\
& + & X_1(\det M - B{\tilde B}-\Lambda^{4M})+X_2(\det N - C{\tilde C}-\Lambda^{4M})
\eeqa

Let us focus on the mesonic branch of the moduli space, i.e. solutions with $\mathcal{B}=\tilde{\mathcal{B}}=\mathcal{C}=\tilde{\mathcal{C}}=0$.

\beq
\begin{array}{lcl}
\partial_{X_1}W=0 & \ \ \Rightarrow \ \ & \det\mathcal{M}=\Lambda^{4M} \\
\partial_{X_2}W=0 & \ \ \Rightarrow \ \ & \det\mathcal{N}=\Lambda^{4M}
\end{array}
\label{deformed_conditions}
\eeq

For simplicity, we concentrate on a particularly simple choice of vev's satisfying \eref{deformed_conditions}

\beq
<\mathcal{M}> =\Lambda^{2} \left[\begin{array}{ccc} 1_{M \times M} & \ &0 \\  0 & & 1_{M \times M} \end{array} \right]
\ \ \ \ \ \            
<\mathcal{N}> =\Lambda^{2} \left[\begin{array}{ccc} 1_{M \times M} & \ & 0 \\ 0 & & 1_{M \times M} \end{array} \right]  
\label{vevs}
\eeq

Denoting $\eta_{ij}$ and $\xi_{ij}$ the fluctuations of $M_{ij}$ and $N_{ij}$ 
around their respective expectation values, and dropping a constant term, 
the superpotential in the Abelian case becomes

\beqa
W & = & \eta_{62} X_{23} \eta_{35} X_{56} - X_{23} X_{35} X_{56} X_{62} - 2\Lambda^{4} - \Lambda^2 (\eta_{63} + \eta_{36} + \eta_{52} +
\eta_{25}) \nonumber \\
& - & \eta_{63} \eta_{36} - \eta_{52} \eta_{25}+\eta_{53} X_{35} + \eta_{26} X_{62} \nonumber \\
&+& X_1(\Lambda^2 (\eta_{63}+ \eta_{52}) + \eta_{63}\eta_{52}-
\eta_{53}\eta_{62})+X_2(\Lambda^2 (\eta_{36}+ \eta_{25}) + \eta_{36}\eta_{25}-\eta_{35}\eta_{26})
\label{W_dP3_fluctuations}
\eeqa

We are interested in looking at energies much smaller than the dynamical scale $\Lambda$. This can be systematically
implemented by taking the large $\Lambda$ limit of the superpotential, which we will call $W'$, and looking
at the approximate equations of motion that follow. For large $\Lambda$, the superpotential becomes

\beq
W' = - \Lambda^2 (\eta_{63}+ \eta_{52}) - \Lambda^2 (\eta_{36}+ \eta_{25}) - \Lambda^2 X_1 (\eta_{63}+ \eta_{52}) - \Lambda^2 X_2 (\eta_{36}+ \eta_{25}) + \mathcal{O}(\Lambda^0)
\eeq

This determines the value of the Lagrange multipliers through

\beqa
\frac{\partial W'}{\partial (\eta_{63}+\eta_{52})}=0 \quad \rightarrow \quad X_1=1 \nonumber \\
\frac{\partial W'}{\partial (\eta_{36}+\eta_{25})}=0 \quad \rightarrow \quad X_2=1 
\eeqa

Plugging this into \eref{W_dP3_fluctuations}, we obtain an expression identical to \eref{W_dP3_mesonic}, with the mesons
replaced by their corresponding fluctuations. The rest of the proof is 
the same as the one in Section \ref{section_dP3_to_conifold}.

This type of discussion makes clear, for example, how the relative minus sign in
the values of the Lagrange multipliers $X_1$ and $X_2$ assumed in \eref{vevs_dP2} is determined.

\section{Description of complex deformations}
\label{toric}

In this section we provide a precise geometric description of the 
complex deformation corresponding to the removal of sub-webs
in the toric diagram of our geometries. For additional details and other examples see \cite{Aganagic:2001ug}.

The basic process in the separation of a sub-web in a toric diagram is the separation of two lines. This basic process
is already present in the complex deformation of the conifold. In 
order to describe it in toric language, recall the
toric data for the conifold
\begin{center}
\begin{tabular}{ccccc}
& $a_1$ & $a_2$ & $b_1$ & $b_2$ \\
Q & $1$ & $1$ & $-1$ & $-1$
\end{tabular}
\end{center}
Namely, one is performing a Kahler quotient of $\IC^4$ by the $U(1)$ action acting on it with the above charges. Physically, the 
conifold is the target of the 2d linear sigma model specified by the above charges for a set of four chiral multiplets. The moment 
map equation (equivalently the D-term equations for the linear sigma model) are
\beqa
|a_1|^2\, + \, |a_2|^2 \, -\, |b_1|^2\, -\, |b_2|^2 \, = \, s
\eeqa
The geometry is toric, namely can be regarded as a fibration of circles over a base. The $U(1)$ action is simply generated by the 
three independent phase rotations of the chiral multiplets, up to the above $U(1)$ action (which is a gauge equivalence).

\begin{figure}[ht]
  \epsfxsize = 8cm
  \centerline{\epsfbox{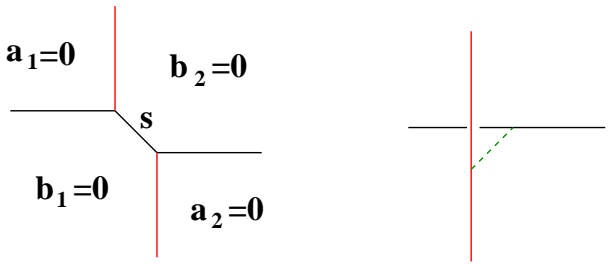}}
  \caption{Toric projection and complex deformation for the conifold.}
  \label{conitori}
\end{figure}

The geometry can be describe using the gauge-invariant quantities $x=a_1a_2$, $y=b_1b_2$, $u=a_1b_1$, $v=a_2b_2$, as the 
hypersurface in $\IC^4$ defined by $xy=uv$. This may be equivalently described by the two equations $xy=z$, $uv=z$. The 
$U(1)$ actions degenerate along lines in the subspace $z=0$. The toric 
projection in Figure \ref{conitori} describes the loci in $z=0$ 
where the $U(1)$ actions degenerate. Notice that $s$ measures the size of the 2-cycle in the resolved conifold.

The complex deformation involving the separation of the two lines, Figure \ref{conitori}b, is possible when $s=0$. To describe it, 
we simply use monomials invariant under the $U(1)$ gauge symmetry associated to $s$, namely $x$, $y$, $u$, $v$, and deform their
constraint to
\beqa
xy-uv=\epsilon
\eeqa
This may be recast as $xy=z+\epsilon$, $uv=z$, showing that there are two different values of $z$ at which the toric fibers 
degenerate. This implies that the two lines have separated from each other.

\medskip

\begin{figure}[ht]
  \epsfxsize = 12cm
  \centerline{\epsfbox{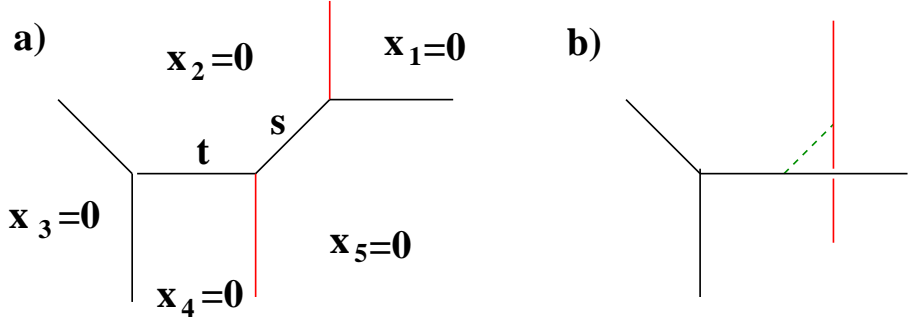}}
  \caption{Toric projection and complex deformation for the SPP.}
  \label{spptori}
\end{figure}

The procedure generalizes to more involved situations. Let us consider the SPP singularity, for which the toric data are 
\begin{center}
\begin{tabular}{cccccc}
      & $x_1$ & $x_2$ & $x_3$ & $x_4$ & $x_5$ \\
$Q_s$ & $1$ & $-1$ & $0$ & $1$ & $-1$\\
$Q_t$ & $0$ & $0$ & $1$ & $-2$ & $1$\\
\end{tabular}
\end{center}
The corresponding D-term equations are
\beqa
|x_1|^2+|x_4|^2 - |x_2|^2-|x_5|^2 =s,\nonumber \\
|x_3|^2+|x_5|^2-2|x_4|^2 = t
\eeqa
There are two parameters $s,t$ which control the size of two independent 2-cycles in the geometry. The toric picture, showing the
degeneration loci of the toric circle actions, is shown in Figure \ref{spptori}a.
The complex structure of the SPP is given by
\beqa
uv = xy^2,
\eeqa
where $x,y,u,v$ are gauge invariant coordinates,
\beqa
x = x_1 x_2,\;\; y = x_3 x_4 x_5, \;\; u= x_1 x_4  x_5^2,\;\; v=x_2 x_3^2 x_4.
\eeqa
The complex structure deformation, in Figure \ref{spptori}b, arises when $s=0$. In order to describe it, we introduce 
variables invariant under $U(1)_s$
\beqa
x=x_1 x_2, \;\; y=x_3x_4x_5\;\; \rho=x_1x_5/x_3\;\; v=x_2x_3^2x_4
\eeqa
(which are well-defined for $x_3\neq 0)$. They satisfy a constraint $x{\tilde y}=\rho {\tilde v}$, which we deform to
\beqa
xy-\rho v=\epsilon
\eeqa
In the complete manifold, using that $\rho=u/y$, we obtain for the complex deformation
\beqa
xy^2\,=\, (\rho v+\epsilon) \, y\, =\, uv \, +\,\epsilon y 
\eeqa
Notice that this geometric argument and the deformed geometry nicely dovetail the field theory argument at the end of section 
\ref{thespp}.

\section{Cones over the $Y^{p,q}$ manifolds}
\label{yp0}

Real cones over the manifolds $Y^{p,q}$ \cite{Gauntlett:2004zh,Gauntlett:2004yd,Gauntlett:2004hh,Gauntlett:2004hs,Martelli:2004wu} provide an infinite family of 6 dimensional singular geometries on which we can place D3-branes. This leads to an infinite class of quiver gauge 
theories, which have been determined in \cite{Benvenuti:2004dy}, and whose 
study is a promising new direction in the gauge/gravity correspondence.

One interesting feature is that the five dimensional $Y^{p,q}$ manifolds have only one collapsing 2-cycle and thus admit a single kind of fractional brane, which triggers a cascading RG flow. Some particular cascades, as well as the KT-like supergravity solutions for the general case, have been 
recently constructed in \cite{Ejaz:2004tr}. The warped throat solutions contain a naked singularity at their tip. A natural question is whether a smooth solution exists, based on a complex deformation of the underlying geometry, and how to understand it from the dual field theory viewpoint.

\begin{figure}[ht]
  \epsfxsize = 10cm
  \centerline{\epsfbox{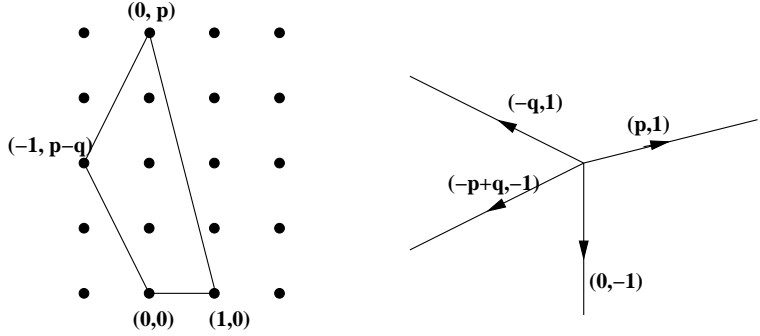}}
  \caption{The toric and web diagram for the cone over the general 
$Y^{p,q}$ manifold. No leg recombination is possible except for the case 
$q=0$.}
  \label{ypq}
\end{figure}

In general these 6 dimensional manifolds correspond to spaces which do not admit 
complex deformations. This can be seen from the web diagrams of those 
spaces, see Figure \ref{ypq}. 

Only in the case of $Y^{p,0}$ a decomposition of the web
into sub-webs is possible. This case is also special, since it corresponds
to a $\IZ_p$ quotient of the conifold. More concretely, defining the
conifold by the equation
\beqa
xy-zw=0
\label{conifold}
\eeqa
the cone over $Y^{p,0}$ is obtained by modding out by the $\IZ_p$ action  
generated by $\theta$, which acts as
\beqa
x\to e^{2\pi i/p}x\quad , \quad
y\to e^{-2\pi i/p}y\quad , \quad
z\to e^{2\pi i/p}z\quad , \quad
w\to e^{-2\pi i/p}w  
\label{quotient1}
\eeqa
which is clearly a symmetry of (\ref{conifold}).

The complex deformation of the manifold is simply the $\IZ_p$ quotient of
the complex deformation of the conifold
\beqa
xy-zw=\epsilon
\label{conifolddef}
\eeqa
The 3-cycle in the deformed space is the Lens space $S^3/\IZ_p$. 

Therefore, although both warped supergravity throats and 
logarithmic RG duality cascades seem to exist for all the $Y^{p,q}$ cases, 
the class of $Y^{p,0}$ manifolds stand out as the only cases which admit 
a complex deformation, presumably removing the infrared singularity of 
their supergravity solutions. Our plan is to center on this class 
and indeed derive the deformation from the viewpoint of the
strong dynamics of the dual gauge theory with fractional branes
in general.

\medskip

For that purpose we need the corresponding quiver gauge theories. These 
can be obtained using the rules in \cite{Benvenuti:2004dy}, but for 
illustration purposes we construct them using their realization as $\IZ_p$ 
quotients of the conifold. This can be done following the ideas in 
\cite{Uranga:1998vf}. Concretely, the conifold theory is $SU(N_1)\times 
SU(N_2)$ 
with fields $A_1$, $A_2$ in the $(\fund,\antifund)$ and $B_1$, $B_2$ in 
the $(\antifund,\fund)$. We also have the superpotential
\beqa
W=A_1B_1A_2B_2-A_1B_2A_2B_1
\eeqa
By the realization of the conifold as the moduli space of the gauge
theory, there is a relation between the fields and the coordinates
$x,y,z,w$. Roughly
\beqa
x\simeq A_1B_1 \quad,\quad
y\simeq A_2B_2 \quad,\quad
z\simeq A_1B_2 \quad,\quad
w\simeq A_2B_1
\eeqa
The action (\ref{quotient1}) can thus be implemented as the action
\beqa
A_1\to e^{2\pi i/p} A_1 \quad , \quad A_2\to e^{-2\pi i/p} A_2 \quad,
\quad B_1\to B_1 \quad , \quad B_2\to B_2
\eeqa
In addition, we have to specify the action of $\theta$ on the $SU(N_1)$
and $SU(N_2)$ Chan-Paton labels. This is done by two order $p$ discrete
gauge transformations, which without loss of generality can be chosen
\beqa
\gamma_{\theta, 1} &=&\diag(\id_{n_0}, e^{2\pi i/p} \id_{n_1},\ldots,
e^{2\pi i(p-1)/p} \id_{n_{p-1}}) \nonumber \\
\gamma_{\theta, 2} &=&\diag(\id_{m_0}, e^{2\pi i/p} \id_{m_1},\ldots,
e^{2\pi i(p-1)/p} \id_{m_{p-1}})
\eeqa
with $\sum_a n_a=N_1$ and $\sum_a m_a=N_2$.

Now we have to project with respect to the combined geometric and
Chan-Paton action. For vector multiplets, the geometric action is trivial,
and we simply get a gauge group
\beqa
SU(n_0)\times \ldots \times SU(n_{p-1})\times
SU(m_0)\times \ldots \times SU(m_{p-1})
\eeqa
while the projection for the chiral multiplets leads to a set of chiral 
multiplets in the following representations
\beqa
(A_1)_{a,a+1} & =  (n_a,\ov{m}_{a+1}) \quad \quad
(A_2)_{a,a-1} & =  (n_a,\ov{m}_{a-1}) \nonumber \\
(B_1)_{a,a} & =  (\ov{n}_a,m_{a}) \quad \quad
(B_2)_{a,a} & =  (\ov{n}_a,m_{a})
\eeqa
The superpotential is directly obtained from the conifold one and reads
\beqa
W=\sum_a\; \left[ (A_1)_{a,a+1} (B_1)_{a+1,a+1} (A_2)_{a+1,a}
(B_2)_{a,a}\, -\, (A_1)_{a,a+1} (B_2)_{a+a,a+1} (A_2)_{a+1,a} (B_1)_{a,a}
\right]\nonumber
\eeqa

The complete result agrees with that using the rules in 
\cite{Benvenuti:2004dy}
(by relabeling $B_{\alpha}\to U^\alpha$, $A_1\to Z$, $A_2\to Y$).
It is easy to check that the quiver for e.g. $Y_{4,0}$ agrees with 
that in figure 8 in \cite{Benvenuti:2004dy}.

\medskip

This gauge theory admits a single kind of fractional brane. The gauge
theory corresponds to $n_a=N+M$, and $m_a=N$. The RG flow presumably leads
to a cascade of Seiberg dualities with structure very similar to that of 
the conifold. Although we have not carried out a complete analysis, we 
would like to
make the following natural proposal. Consider all the nodes $SU(N)$ to have
equal coupling at some UV scale, and all nodes $SU(N+M)$ to have equal
coupling. Namely, we consider the couplings to respect the $\IZ_p$ 
symmetry
of the quiver. As we run to the IR, the nodes $SU(N+M)$ get to strong
coupling. Let us Seiberg dualize them simultaneously (to do it in
practice, we may do it sequentially, but presumably the order is not
important). After this, we obtain a similar quiver, with all ranks $N+M$
replaced by $N-M$. So next one should dualize all the nodes of rank $N$,  
etc. This just amounts to inheriting the cascade from the parent to the 
orbifold theory.

Let us now consider the infrared behavior of the cascade. For $N$ a
multiple of $M$ (in which case we center in what follows) the endpoint of
the cascade is a theory of $p$ decoupled $N=1$ SYM nodes, with equal
gauge coupling (or dynamical scale) due to the $\IZ_p$ symmetry of the 
flow. The unique dynamical scale should be associated with a finite-size 
3-cycle in a deformed geometry.

In order to check that the geometry at the tip of the throat is the  
deformed geometry described above, we consider the gauge theory
describing the dynamics of $M$ D3-brane probing the IR theory. Namely, 
using the by now familiar technique we take the quiver theory with group
\beqa
\prod_a SU(2M)_a\times \prod_a SU(M)_a
\eeqa
The nodes $SU(2M)_a$ condense, so we introduce the mesons
\beqa
{\cal M}=\left[ \begin{array}{cc} M_{a,a+1} & {\tilde M}_{a,a+1} \cr
M_{a,a-1} & {\tilde M}_{a,a-1}\end{array} \right] =
\left[ \begin{array}{cc} (A_1)_{a,a+1}(B_1)_{a+1,a+1} & (A_1)_{a,a+1}
(B_2)_{a+1,a+1}\cr (A_2)_{a,a-1} (B_1)_{a-1,a-1} & (A_2)_{a,a-1}
(B_2)_{a-1,a-1}
\end{array}
\right]
\eeqa
In terms of these, the superpotential reads
\beqa
W = \sum_a\; \left[ M_{a,a+1} {\tilde M}_{a+1,a} - {\tilde M}_{a,a+1} M_{a+1,a} \right]
\eeqa
We now should impose the quantum constraint, and pick the mesonic branch.
Along the mesonic branch, all the $SU(M)_a$ gauge groups are broken to a 
single diagonal combination. Therefore all mesons transform in
the adjoint representation of this gauge group. Imposing the constraint as
a superpotential and centering in the Abelian case as usual, we have
\beqa
W = \sum_a\; \left[ M_{a,a+1} {\tilde M}_{a+1,a} - {\tilde M}_{a,a+1} M_{a+1,a}
- M_{a,a+1} {\tilde M}_{a,a-1} + M_{a,a-1}{\tilde M}_{a,a+1} \right]
\eeqa
Notice that we have $4pM^2$ meson degrees of freedom. However, they have
to satisfy the F-term equations
\beqa
{\tilde M}_{a+1,a}& ={\tilde M}_{a,a-1} \quad \quad
M_{a+1,a}& =M_{a,a-1} \nonumber \\
{\tilde M}_{a,a+1}& ={\tilde M}_{a-1,a} \quad \quad
M_{a,a+1}& =M_{a-1,a}
\eeqa
These are apparently $4pM^2$ relations. However, they are not all
independent. This can be seen by noticing that they only fix the relative
vevs of the mesons for different values of $a$, but they do not fix the
overall size of a given kind of meson. Therefore, there are four operators
whose vevs are not fixed by the above conditions. They are

\beqa
M_{11}=\prod_a M_{a,a+1} \quad, \quad M_{12}= \prod_a {\tilde M}_{a,a+1}   
\quad , \quad
M_{21}=\prod_a M_{a+1,a} \quad, \quad M_{22}= \prod_a {\tilde M}_{a+1,a}
\label{relp}
\eeqa

Notice however that the original mesons are also constrained by the
quantum constraint (which is obtained from $\partial W/\partial X_a=0$
before going into the mesonic branch etc). This implies that the final
operators have to satisfy

\beqa
M_{11} M_{22} - M_{12}M_{21} =\Lambda^{P}
\eeqa

This moduli space indeed corresponds to a deformed space. Moreover, the  
fact that the fundamental mesons are related to the above fields by the
order $p$ relation (\ref{relp}) shows that the final space is a
$\IZ_p$ quotient of the deformed conifold.

Hence the whole family of $Y^{p,0}$ cones is closely related to the 
KS conifold, and a generalization of the complex cone over $F_0$, which 
is the case $p=2$ in the above language. The field theory argument plus 
the geometric analysis strongly support the existence of a smooth 
supergravity solution describing a complete RG flow for these theories. 
Indeed, these exist and are given simply by the $\IZ_p$ quotient of the 
KS solution.


\bibliographystyle{JHEP}

\end{document}